\documentclass[prd,preprint,tightenlines,showpacs,nofootinbib,eqsecnum,amsfonts,amssymb,amsmath]{revtex4}

\newcommand{\be}{\begin{equation}}
\newcommand{\ee}{\end{equation}}
\newcommand{\bea}{\begin{eqnarray}}
\newcommand{\eea}{\end{eqnarray}}

\newcommand{\dbi}{\partial_{i}}
\newcommand{\dbj}{\partial_{j}}
\newcommand{\dhi}{\partial^{i}}

\newcommand{\dbk}{\partial_{k}}
\newcommand{\dbl}{\partial_{l}}
\newcommand{\dhk}{\partial^{k}}

\def\dd{{\rm d}}

\def\HH{\mathcal{H}}
\newcommand{\Christoffel}[3]{\Gamma^{#1}_{#2 #3}}

\def\sva{\sigma_{_{\rm V}a}}
\def\svb{\sigma_{_{\rm V}b}}
\def\stl{\sigma_{_{\rm T}\lambda}}
\def\spar{\sigma_{_\parallel}}
\def\stplus{\sigma_{_{\rm T}+}}
\def\stcross{\sigma_{_{\rm T}\times}}
\newcommand{\urel}[1]{{\tt #1}}

\begin{document}

\title{Theory of cosmological perturbations in an anisotropic universe}

\author{Thiago S. Pereira}
 \email{thiago@if.usp.br}
 \altaffiliation{also at
             Institut d'Astrophysique de Paris,
             Universit\'e Pierre~\&~Marie Curie - Paris VI,
             CNRS-UMR 7095, 98 bis, Bd Arago, 75014 Paris, France.}
 \affiliation{
             Instituto de F\'isica,
             Universidade de S\~ao Paulo
             CP 66318, 05315-970 S\~ao Paulo, Brazil.}

\author{Cyril Pitrou}
 \email{pitrou@iap.fr}
 \affiliation{
             Institut d'Astrophysique de Paris,
             Universit\'e Pierre~\&~Marie Curie - Paris VI,
             CNRS-UMR 7095, 98 bis, Bd Arago, 75014 Paris, France.}

\author{Jean-Philippe Uzan}
 \email{uzan@iap.fr}
 \affiliation{
             Institut d'Astrophysique de Paris,
             Universit\'e Pierre~\&~Marie Curie - Paris VI,
             CNRS-UMR 7095, 98 bis, Bd Arago, 75014 Paris, France.}

\begin{abstract}
This article describes the theory of cosmological perturbations around a homogeneous and
anisotropic universe of the Bianchi~$I$ type. Starting from a general parameterisation of the
perturbed spacetime {\it \`a la Bardeen}, a complete set of gauge invariant variables is
constructed. Three physical degrees of freedom are identified and it is shown that, in the case
where matter is described by a scalar field, they generalize the Mukhanov-Sasaki variables. In
order to show that they are canonical variables, the action for the cosmological perturbations at
second order is derived. Two major physical imprints of the primordial anisotropy are identified:
(1) a scalar-tensor ``see-saw'' mechanism arising from the fact that scalar, vector and tensor
modes do not decouple and (2) an explicit dependence of the statistical properties of the density
perturbations and gravity waves on the wave-vector instead of its norm. This analysis extends, but
also sheds some light on, the quantization procedure that was developed under the assumption of a
Friedmann-Lema\^{\i}tre background spacetime, and allows to investigate the robustness of the
predictions of the standard inflationary scenario with respect to the hypothesis on the symmetries
of the background spacetime. These effects of a primordial anisotropy may be related to some
anomalies of the cosmic microwave background anisotropies on large angular scales.
\end{abstract}
 \date{5 July 2007}
 \maketitle

Inflation~\cite{lindebook,pubook} (see Ref.~\cite{linde2007} for a recent review of its status and
links with high energy physics) is now a cornerstone of the standard cosmological model. Besides
solving the standard problems of the big-bang model (homogeneity, horizon, isotropy,
flatness,...), it provides a scenario  for the origin of the large scale structure of the
universe. In its simplest form, inflation has very definite predictions: the existence of
adiabatic initial scalar perturbations and gravitational waves, both with Gaussian statistics and
an almost scale invariant power spectrum~\cite{ref:inf,mbf}. Other variants, which in general
involve more fields, allow e.g. for isocurvature perturbations~\cite{iso},
non-Gaussianity~\cite{ng}, and modulated fluctuations~\cite{bku}. All these features let us hope
that future data will allow a better understanding of the details (and physics) of this primordial phase.

The predictions of inflation are in agreement with most cosmological data and in particular those
of the cosmic microwave background (CMB) by the WMAP satellite~\cite{wmap3}. The origin of the
density perturbations is related to the amplification of vacuum quantum fluctuations of a scalar
field during inflation. In particular, the identification of the degrees of freedom that should be
quantized (known as the Mukhanov-Sasaki variables~\cite{MSvar}), has been performed assuming a
Friedmann-Lema\^{\i}tre background spacetime~\cite{mbf}. This means that homogeneity and isotropy
(and even flatness) are in fact assumed from the start of the computation. In the standard lore,
one assumes that inflation lasts long enough so that all classical inhomogeneities (mainly spatial
curvature and shear) have decayed so that it is perfectly justified to start with a flat
Friedmann-Lema\^{\i}tre background spacetime when dealing with the computation of the primordial
power spectra for the cosmologically observable modes. This is backed up by the ideas of chaotic
inflation and eternal inflation~\cite{linde2007}. Note however that a (even small) deviation from
flatness~\cite{inflK} or isotropy~\cite{infS} may have an impact on the dynamics of inflation. It
would however be more satisfactory to start from an arbitrary spacetime and understand (1) under
which conditions it can be driven toward a Friedmann-Lema\^{\i}tre spacetime during inflation and
(2) what are the effects on the evolution and quantization of the perturbations.

The first issue has  been adressed by considering the onset of inflation in inhomogenous and
spherically symmetric universes, both numerically in Ref.~\cite{inhomo_numeric} and
semi-analytically in Ref.~\cite{inhomo_analytic}. The isotropization of the universe was also
investigated by considering the evolution of four-dimensional Bianchi
spacetimes~\cite{bianchiold,bianchiinf0,bianchiinf1} and even Bianchi braneworld~\cite{braneinf}.
No study has focused on the second issue, i.e. the evolution and the quantization of perturbations
during a non-Friedmannian inflationary stage, even though the quantization of test fields and
particle production in anisotropic spacetime has been considered~\cite{Qaniso}. Such an analysis
would shed some light on the specificity of the standard quantization procedure which assumes a
flat Friedmannian background (see however Ref.~\cite{vCurv}).

From an observational perspective, a debate concerning possible anomalies on large angular scales
in the WMAP has recently driven a lot of activity. Among these anomalies, we count the lack of
power in the lowest multipoles, the alignment of the lowest multipoles, and an asymmetry between
the two hemispheres (see e.g. Refs.~\cite{lowQ}). The last two, which point toward a departure
from the expected statistical isotropy of the CMB temperature field, appear much stronger. Various
explanations for these anomalies, besides an understood systematic effect that may be related to
foreground (see e.g. Ref.~\cite{prunet}), have been proposed (such as e.g. the imprint of the
topology of space~\cite{topologie,modetopo} the breakdown of local isotropy due to multiple scalar
fields~\cite{picon} or the existence of a primordial preferred
direction~\cite{CMB_bianchi,acker}).

The broken statistical isotropy of the temperature fluctuations may also be related to a violation
of local isotropy, and thus from a departure from the Friedmann-Lema\^{\i}tre symmetries. This can
arise either from a late time evolution of the universe (see e.g.
Refs.~\cite{Jaffe2005,Jaffe2006,polbianchi} in which it is argued that the subtraction of a
Bianchi~$VII_h$ leaves a statistically isotropic CMB sky) or from the primordial dynamics which
would have imprinted the broken statistical isotropy in the initial conditions. The latter has
recently been advocated on the basis of a cylindrically symmetric Bianchi~$I$ inflationary
model~\cite{CMB_bianchi}. In these models, the shear decays as the inverse of the third power of
the scale factor so that it can play a significant role only in the early stage of the
inflationary period. Isotropy is asymptotically reached during inflation and the whole subsequent
cosmological evolution can be approximated by a Friedmann-Lema\^{\i}tre universe. It follows (1)
that the anisotropy is only imprinted in the largest wavelengths and (2) that the constraints on
the shear of the observable universe from the isotropy of the
CMB~\cite{LimitShear1,LimitShear2,LimitShear3,LimitShear4} or big-bang nucleosynthesis~\cite{bbn}
are satisfied.

The primordial Bianchi~$I$ phase modifies the evolution of the modes (in particular gravity wave
and scalar perturbations shall be coupled through the shear) and initial conditions (and thus the
quantization procedure) has to be performed in a consistent way with the symmetries of the
background spacetime during inflation (see however Ref.~\cite{CMB_bianchi} for a proposal in a
locally rotational
invariant and homogeneous spacetime of the Kantowski-Sachs family).\\

In this article, we investigate the general theory of gauge invariant perturbations about a
Bianchi~$I$ background spacetime during inflation. Bianchi universes are spatially homogeneous
spacetimes and are thus of first importance in cosmology since they express mathematically the
cosmological principle. The study of perturbations in Bianchi~$I$ was roughed out in
Ref.~\cite{TomitaDen} where the Bardeen formalism was used (see also Ref.~\cite{NohHwang} and
Ref.~\cite{Abbotetal} for the case of higher-dimensional Kaluza-Klein models). A similar work was
undertaken in the $1+3$ covariant formalism~\cite{Dunsby} but the identification of gravitational
waves and the quantization procedure was not adressed (see Ref.~\cite{pitrou07} for the
generalisation of the Mukhanov-Sasaki variables in this formalism).

Thus, starting from a general parameterisation of the perturbed spacetime {\it \`a la
Bardeen}~\cite{Bardeen}, we will define in Section~\ref{sec3}, a scalar-vector-tensor
decomposition and construct gauge invariant variables. Contrary to the Friedmann-Lema\^{\i}tre
case, these three types of perturbations will be coupled through the shear. In section~\ref{sec4},
we derive the perturbation equations. We then show in Section~\ref{sec5} that they can be reduced
to a set of coupled reduced equations with a mixing between scalar and tensor modes; special care
will be taken to vector modes. This work will allow to generalize the Mukhanov-Sasaki variables and
paves the way to the study of the cosmological signatures of a primordial anisotropy~\cite{pdu}.

\section{Cosmological dynamics of Bianchi~$I$ universes}\label{sec2}

Bianchi spacetimes enjoy a group of isometries simply transitive on spacelike hypersurfaces (see
Refs.~\cite{vElst,BianchiMath1,BianchiMath2} for a mathematical expositions on Bianchi
spacetimes). Thus, they are homogeneous. It follows that the cosmic time $t$ is the only essential
dynamical coordinate and Einstein equations will reduce to ordinary differential equations. The
dimension of their group of isotropy~\cite{vElst}, that is the group of isometries leaving a given
point fixed, is $q=0$.

\subsection{General form of the metric}

Bianchi~$I$ spacetimes are the simplest anisotropic universe models. They allow for different
expansion factors in three orthogonal directions. In comoving coordinates, the metric takes the
general form
\begin{equation}\label{metric1}
 \dd s^2 = g_{\mu\nu}\dd x^\mu\dd x^\nu = -\dd t^2+ \sum_{i=1}^3 X_i^2(t)\,\left(\dd x^i\right)^2\,.
\end{equation}
It includes Friedmann-Lema\^{\i}tre spacetimes as a subcase when the three scale factors are
equal. The average scale factor, defined by
\begin{equation}\label{def_a}
 a(t) \equiv \left[X_1(t)X_2(t)X_3(t)\right]^{1/3}\,,
\end{equation}
characterizes the volume expansion. It follows that we can recast the metric~(\ref{metric1}) as
\begin{equation}\label{metric2}
 \dd s^2 = -\dd t^2+ a^2(t)\gamma_{ij}(t)\dd x^i\dd x^j\,.
\end{equation}
The ``spatial metric'' $\gamma_{ij}$ is the metric on constant time hypersurfaces. It can be
decomposed as
\begin{equation}\label{metricdec}
 \gamma_{ij}= \hbox{exp}\left[{2\beta_i(t)}\right]\delta_{ij}\,,
\end{equation}
with the constraints
\begin{equation}\label{beta}
 \sum_{i=1}^{3}\beta_{i}=0\,.
\end{equation}
Let us emphasize that $\beta_i$ are not the components of a vector so that they are not subjected
to the Einstein summation rule. Note also that all latin indices $i,j,\ldots$ are lowered with the
metric $\gamma_{ij}$. The decomposition~(\ref{metricdec}) implies that $\dot\gamma_{ij} =
2\dot\beta_i\gamma_{ij}$, where a dot refers to a derivative with respect to the cosmic time, and
it can be checked that the spatial hypersurfaces are flat. This relation, together with the
constraint~(\ref{beta}), implies that the determinant of the spatial metric is constant
$$
 \dot\gamma = \gamma^{ij}\dot{\gamma}_{ij} = 0\,.
$$
This simply means that any comoving volume remains constant during the expansion of the universe,
even if this expansion is anisotropic. We define the shear as
\begin{equation}
 \hat\sigma_{ij}\equiv\frac{1}{2}\dot\gamma_{ij}
\end{equation}
and introduce the scalar $\hat\sigma^2\equiv\hat\sigma_{ij}\hat\sigma^{ij}$. This definition is
justified from the relation to the $1+3$ covariant formalism (see Appendix~\ref{appA3}). Let us
emphasize at this point that $(\gamma^{ij})^\cdot=-2\hat\sigma^{ij}$ differs from $\dot{\gamma}^{ij}
\equiv\gamma^{ip} \gamma^{jk}\dot{\gamma}_{pk}=+2\hat\sigma^{ij}$.

Introducing the conformal time as $\dd t \equiv a \dd\eta$, the metric~(\ref{metric2}) can be
recast as
\begin{equation}\label{metricconforme}
 \dd s^2 = a^2(\eta)\left[-\dd \eta^2+ \gamma_{ij}(\eta)\dd x^i\dd x^j\right].
\end{equation}
We define the comoving Hubble parameter by $\mathcal{H}\equiv a'/a$, where a prime refers to a
derivative with respect to the conformal time. The shear tensor, now defined as
\begin{equation}\label{defsigmaij}
 \sigma_{ij} \equiv \frac{1}{2}\gamma_{ij}',
\end{equation}
is related to $\hat\sigma_{ij}$ by $\sigma_{ij}=a\hat\sigma_{ij}$. From the relation
$(\gamma_{ij})'=\gamma^\prime_{ij}=2\beta^\prime_i\gamma_{ij}$, the definition
\begin{equation}
 \sigma^2\equiv \sigma_{ij}\sigma^{ij}
\end{equation}
is explicitely given by
\begin{equation}\label{beta2}
 \sigma^2=\sum_{i=1}^3(\beta_i')^2\,,
\end{equation}
and is related to its cosmic time analogous by $\sigma=\hat\sigma a$. Again, we stress that
$(\gamma^{ij})'=-2\sigma^{ij}$ differs from $({\gamma}')^{ij} \equiv\gamma^{ip}
\gamma^{jk}\gamma'_{pk}=+2\sigma^{ij}$.

\subsection{Background equations}

We concentrate on an inflationary phase during which the matter content of the universe is assumed
to be described by a minimally coupled scalar field, $\varphi$, with stress-energy tensor
\begin{equation}
 T_{\mu\nu}=\partial_\mu\varphi\partial_\nu\varphi-\left(\frac{1}{2}\partial_\alpha\varphi
 \partial^\alpha\varphi + V\right)g_{\mu\nu}\,.
\end{equation}
Making use of the expressions (\ref{appg00}-\ref{appgij}) (see Appendix~\ref{appA2}), we easily
obtain the Friedmann equations
\begin{eqnarray}
 &&\HH^2 = \frac{\kappa}{3}\left[\frac{1}{2}\varphi^{\prime2}+ V(\varphi) a^2 \right] +
 \frac{1}{6}\sigma^2,\label{e:1.12}\\
 &&\HH'= -\frac{\kappa}{3}[\varphi^{\prime2} - V(\varphi)a^2] -\frac{1}{3}\sigma^2\label{e:1.13}\\
 &&(\sigma^i_j)' = - 2 \HH \sigma^i_j\,,\label{e:1.14}
\end{eqnarray}
where $\kappa\equiv 8\pi G$. The first two are similar to the ones usually used in a
Friedmann-Lema\^{\i}tre universe, up to the contribution of the shear (which acts as an extra
massless field). The latter equation arises from the trace-free part of the ``$ij$''-Einstein
equations and gives an extra-equation compared to the Friedmann-Lema\^{\i}tre case. We can easily
integrate it and conclude that the shear evolves as
\begin{equation}
 \sigma^i_j=\frac{{\cal S}^i_j}{a^2}
\end{equation}
where ${\cal S}^i_j$ is a constant tensor, $({\cal S}^i_j)'=0$. This implies that
\begin{equation}\label{SigK}
 \sigma^2=\frac{{\cal S}^2}{a^4}\Rightarrow\hat\sigma^2=\frac{{\cal S}^2}{a^6}\,,
\end{equation}
(with ${\cal S}^2\equiv {\cal S}^i_j {\cal S}^j_i$) from which we deduce that
\begin{equation}\label{SigKprim}
 \sigma'=-2\HH\sigma\,.
\end{equation}
Let us note that these equations can be combined to give
\begin{equation}\label{e:a0}
 2\HH^2 + \HH'=\kappa a^2 V\,,
 \qquad
 \kappa(\varphi')^2 = 2 \HH^2 - 2 \HH'- \sigma^2\,.
\end{equation}
These equations are completed by a Klein-Gordon equation, which keeps its Friedmann-Lema\^{\i}tre
form,
\begin{equation}
 \varphi''+2\HH\varphi'+a^2V_\varphi=0.
\end{equation}
The general solution for the evolution of the scale factor from these equations is detailed in
Appendix~\ref{appA4}.

\section{Gauge invariant variables}\label{sec3}

This section is devoted to the definition of the gauge invariant variables that describe the
perturbed spacetime. We follow a method {\it \`a la Bardeen}. In order to define scalar, vector
and tensor modes, we will need to use a Fourier transform. We start, in \S~\ref{sec3_1}, by
recalling its definition and stressing its differences with the standard Friedmann-Lema\^{\i}tre
case. In \S~\ref{sec3_2}, we perform a general gauge transformation to identify the gauge
invariant variables.

\subsection{Mode decomposition}\label{sec3_1}

\subsubsection{Definition of the Fourier transform}

We decompose any quantity in Fourier modes as follows. First, we pick up a comoving coordinates
system, $\{x^i\}$, on the constant time hypersurfaces. Then, we decompose any scalar function as
\begin{equation}\label{fft}
 f\left(x^j,\eta\right)= \int\frac{\dd^3k_i}{\left(2\pi\right)^{3/2}}
         \,\hat f\left(k_i,\eta\right)\,\hbox{e}^{\mathrm{i}k_ix^i}\,,
\end{equation}
with the inverse Fourier transform
\begin{equation}\label{ffti}
 \hat f\left(k_j,\eta\right)= \int\frac{\dd^3x^i}{\left(2\pi\right)^{3/2}}
         \, f\left(x^i,\eta\right)\, \hbox{e}^{-\mathrm{i}k_ix^i}\,.
\end{equation}
In the Fourier space, the comoving wave co-vectors $k_i$ are constant, $k_i'=0$. We now define
$k^i\equiv\gamma^{ij}k_j$ that is obviously a time-dependent quantity. Contrary to the standard
Friedmann-Lema\^{\i}tre case, we must be careful not to trivially identify $k_i$ and $k^i$, since
this does not commute with the time evolution. Note however that $x_ik^i=x^ik_i$ remains constant
so that there is no extra-time dependency entering our definitions~(\ref{fft}-\ref{ffti}). In the
following of this article, we will forget the ``hat'' and use the notation
$f\left(x^j,\eta\right)$ and $f\left(k_j,\eta\right)$ both for a function and its Fourier
transform.

It is easily checked, using the definition~(\ref{defsigmaij}), that
\begin{equation}
 (k^i)'=-2\sigma^{ip}k_p\,.
\end{equation}
This implies that the modulus of the comoving wave vector, $k^2 =  k^{i}k_{i} =
\gamma^{ij}k_{i}k_{j}$, is now time-dependent and that its rate of change is explicitely given by
\begin{equation}
 \frac{k'}{k} = - \sigma^{ij}\hat k_{i}\hat k_{j}\,,
\end{equation}
where we have introduced the unit vector
\begin{equation}
 \hat k_i \equiv \frac{k_i}{k}.
\end{equation}
This vector will turn to be particularly useful for our analysis and we note that it evolves as
\begin{equation}
 (\hat k^i)'= (\sigma^{pq}\hat k_p\hat k_q) \hat k^i - 2\sigma^{ip}\hat k_p\,.
\end{equation}
Indeed, we find that in the standard Friedmann-Lema\^{\i}tre limit ($\sigma_{ij}=0$), $k^i$ and
$k$ are constant.

\subsubsection{Decomposition of the vector and tensor modes}\label{sec_mode_dec}

We shall now decompose the perturbations in their scalar, vector and tensor modes.

Any (3-dimensional) vector field, $V^i$, can be decomposed as
\begin{equation}
 V_i = \partial_i V + \bar V_i\,,
 \qquad\hbox{with}\qquad
 \partial^i\bar V_i=0\,,
\end{equation}
Note that we have chosen orthogonal (but not Cartesian) coordinates on
the (Euclidean) spatial sections (in particular spatial flatness and
homogeneity imply that in these coordinates the Christoffel symbols
vanish and that $\partial_k\gamma_{ij}=0$). It follows that its Fourier
components can be split as
\begin{equation}
 V_i = k_i V + \bar V_i,
 \qquad\hbox{with}\qquad
 k^i\bar V_i=0,
\end{equation}
so that $\bar V^i$ lives in the subspace perpendicular to $k^i$. This is a 2-dimensional subspace
so that $V_i$ has been split into 1 scalar ($V$) and two vector modes ($\bar V_i$) that correspond
to transverse modes. Let us now consider the base $\{e^1,e^2\}$ of the subspace perpendicular to
$k^i$. By construction, it satisfies the orthonormalisation conditions
$$
 e^a_i k_j \gamma^{ij}=0,\quad
 e^a_i e^b_j \gamma^{ij}= \delta^{ab}.
$$
Such a basis is defined up to a rotation about the axis $k^i$. Now, the vector modes
can be decomposed on this basis as
\begin{equation}\label{dec_vector}
 \bar V_i(k_i,\eta)=\sum_{a=1,2}V_a(\hat k_i,\eta) \, e_i^a(\hat k_i)\, ,
\end{equation}
which defines the two degrees of freedom, $V_a$, which depend on $\hat k^i$ since the
decomposition differs for each wave number. The two basis vectors allow to define a projection
operator onto the subspace perpendicular to $k^i$ as
\begin{equation}\label{P}
 P_{ij} \equiv e_i^1e_j^1 + e_i^2e_j^2 =\gamma_{ij} -\hat k_i\hat k_j.
\end{equation}
It trivially satisfies $P^i_jP^j_k=P^i_k$, $P^i_jk^j=0$ and $P^{ij}\gamma_{ij}=2$. It is also the
projector on vector modes so that we can always make the scalar-vector decomposition
\begin{eqnarray}
 V_i   &=& [\hat k^jV_j]\hat k_i + P^j_iV_j\, .
\end{eqnarray}

Analogously, any (3-dimensional) symmetric tensor field, $V_{ij}$, can be decomposed as
\begin{equation}\label{SVTdec}
 V_{ij} = T\gamma_{ij}+ \Delta_{ij}S
        +2\partial_{(i}\bar V_{j)}
        +2\bar V_{ij}\,,
\end{equation}
where $\Delta_{ij}\equiv \partial_i\partial_j-\Delta \gamma_{ij}/3$ and
\begin{equation}
 \partial_i \bar V^i=0, \quad
 \bar V_i^i=0=\partial_i\bar V^{ij}.
\end{equation}
The symmetric tensor $\bar V_{ij}$ is transverse and trace-free. Hence it has only two independent
components and can be decomposed as
\begin{equation}\label{dec_tensor}
 \bar V_{ij}(k_i,\eta)=\sum_{\lambda=+,\times} V_\lambda(k^i,\eta)
 \, \varepsilon_{ij}^\lambda(\hat k_i)
\end{equation}
where the polarization tensors have been defined as
\begin{equation}\label{defespilonij}
 \varepsilon_{ij}^\lambda=\frac{e_i^1e_j^1 - e_i^2e_j^2}{\sqrt{2}}\delta^\lambda_+
 + \frac{e_i^1e_j^2 + e_i^2e_j^1}{\sqrt{2}}\delta^\lambda_\times.
\end{equation}
It can be checked that they are traceless ($\varepsilon_{ij}^\lambda \gamma^{ij}=0$), transverse
($\varepsilon_{ij}^\lambda k^i=0$), and that the two polarizations are perpendicular
($\varepsilon_{ij}^\lambda \varepsilon^{ij}_{\mu}=\delta^\lambda_{\mu}$). This defines the two
tensor degrees of freedom.

In order to deal with the properties of the polarization tensors, it is useful to define two new
quantities
\begin{equation}
 Q_{ij} \equiv e_i^1e_j^2-e_i^2e_j^1\,,\qquad \hbox{and}\qquad
\eta_{\lambda\mu}\equiv\delta^+_\lambda\delta^\times_{\mu} - \delta^+_\mu\delta^\times_{\lambda}\,.
\end{equation}
The tensor $Q_{ij}$ trivially satisfies
\begin{equation}\label{pq}
    P_{ij}Q^{ij}=0\,,
    \qquad
    Q_{ij}Q^{ij}=2\,.
\end{equation}
They allow us to simplify the product of two and three polarization tensors as
\begin{equation}\label{epsilon2}
 \varepsilon_{ik}^\lambda\varepsilon^{k\mu}_j=\frac{1}{2}\left(P_{ij}
 \delta^{\lambda\mu}+Q_{ij} \eta^{\lambda\mu}\right)\,,
 \qquad
 \varepsilon_{ik}^\lambda\varepsilon^{kj}_{\mu}\varepsilon^{i}_{j\nu} = 0 \,.
\end{equation}
Introducing the projector operator on tensor modes by
$$
 \Lambda^{ab}_{ij}=P^a_iP^b_j-\frac{1}{2}P_{ij}P^{ab}\,,
$$
and the ``trace extracting'' operator
$$
T_{i}^{j}=\hat k_i \hat k^j-\frac{1}{3}\delta^j_i\, ,
$$
the scalar-vector-tensor terms in the decomposition of Eq.~(\ref{SVTdec}) are
extracted as follows
\begin{eqnarray}
 V_{ij}&=& \left[\frac{1}{3}V_{ab}\gamma^{ab}\right]\gamma_{ij}
         + \left[\frac{3}{2}V_{ab}T^{ab}\right]T_{ij}
         + 2\hat k_{(i}\left[P^a_{j)}\hat k^bV_{ab}\right]
         +\Lambda^{ab}_{ij}V_{ab}.
\end{eqnarray}
In this expression, $V_{ij}$ has been split into 2 scalars ($T$ and $S$), two vector modes
($\bar V_i$) and two tensor modes ($\bar V_{ij}$). Thus, we can always split any equation $V_i=0$
by projecting along $\hat k^i$ (scalar) and $P^i_j$
(vector) and any equation $V_{ij}=0$ by projecting along $\gamma^{ij}$ (scalar), $T^{ij}$
(scalar), $P^i_l\hat k^j$ (vector) and $\Lambda_{ab}^{ij}$ (tensor).

\subsubsection{Properties of the projectors, polarization vectors and tensors}\label{sec_time_pol}

The previous SVT decomposition matches the one used in the perturbation theory about a
Friedmann-Lema\^ {\i}tre spacetime. There is however an important difference that we will have to
deal with. As we pointed out, in a Bianchi~$I$ spacetime, the spatial metric is time-dependent. It implies in
particular that, in order to remain an orthonormal basis perpendicular to $k^i$ during the time evolution, the
polarization vectors, and thus the polarization tensors, must have a non-vanishing time
derivative. Indeed, since $(k_i)'=0$, the vector $(e_a^i)'$ is orthogonal to $k_i$ and is thus a linear
combination of $e^1$ and $e^2$, that is
$$
 (e^i_a)'= \sum_b \mathcal{R}_{ab}\, e^i_b\,.
$$
In each time hypersurface, there is a remaining freedom in the choice of this basis because of the
rotational invariance around $k^i$. We can continuously fix the choice of the basis by imposing
$$
 \mathcal{R}_{[ab]}=0\,.
$$
The orthonormalisation condition implies that $(e_a^ie^b_i)'=0$ and thus that
\begin{equation}
 \mathcal{R}_{ab}= -\sigma_{ij}\,e^i_a e^j_b.
\end{equation}
Consequently, the time derivative of any polarization vector is given by
\begin{equation}\label{e:eaprim}
 (e^a_i)' = \sum_b \mathcal{R}_{ab}\, e_i^b + 2\sigma_{ij} e_a^j\,,
\end{equation}
from which we deduce
\begin{equation}
 k^i(e^a_i)' = 2\sigma^{pi}\,k_p\,e^a_i\,.
\end{equation}

This allows us to derive the expression of the time derivative of the polarization tensor. Starting
from their definitions~(\ref{defespilonij}), we easily obtain that
\begin{equation}\label{eprim}
 \left(\varepsilon_{ij}^{\lambda}\right)'
     = -(\sigma^{kl}\varepsilon_{kl}^\lambda) P_{ij}
                                  -(\sigma^{kl}P_{kl}) \varepsilon_{ij}^{\lambda}
                                  +4\sigma_{(i}^k\varepsilon_{j)k}^{\lambda}\,,
\end{equation}
from which we can deduce some useful algebra
\begin{equation}
 k^ik^j\left(\varepsilon_{ij}^{\lambda}\right)'=0\,,
 \quad
 \gamma^{ij}\left(\varepsilon_{ij}^{\lambda}\right)'=2\sigma^{ij}\varepsilon_{ij}^\lambda\,,
 \quad
 k^i\left(\varepsilon_{ij}^{\lambda}\right)'=2\sigma^{ip}k_p\varepsilon_{ij}^\lambda\,.
\end{equation}
We also have that
\begin{equation}
  \left(\varepsilon_{j}^{i\lambda}\right)'\,\varepsilon_{i}^{j\mu} = 0\,.
\end{equation}
We gather in Appendix~\ref{appB} some other useful relations concerning the polarization vectors
and tensors.

For the sake of completeness, we shall define here two important matrices for the following of our
computation,
\begin{equation}\label{defMab}
 {\cal M}_{ab}^\lambda \equiv \varepsilon^\lambda_{ij}e^i_a e^j_b\,,
\end{equation}
which is manifestly symmetric in $ab$ and
\begin{equation}\label{defNab}
 {\cal N}_{ab} \equiv Q_{ij}e^i_a e^j_b\,,
\end{equation}
which is anti-symmetric in $ab$. We stress that $a$ and $\lambda$ are not indices but only labels.
It can easily be checked that
\begin{equation}
 {\cal M}_{ab}^\lambda = \frac{1}{\sqrt{2}}
      \left(\begin{array}{cc}
             1 &0\\0&-1
             \end{array}\right)\delta^\lambda_+
             +
     \frac{1}{\sqrt{2}}
      \left(\begin{array}{cc}
             0 &1\\1&0
             \end{array}\right)\delta^\lambda_\times\,,
\end{equation}
and that
\begin{equation}
 \sum_a {\cal M}^\lambda_{aa} = 0\,.
\end{equation}

\subsection{Defining gauge invariant variables}\label{sec3_2}

\subsubsection{Gauge invariant variables for the geometry}

Let us consider the most general metric of an almost Bianchi~$I$ spacetime. It can always be
decomposed as
\begin{equation}\label{dmet1}
  \dd s^{2}=a^{2}\left[-\left(1+2A\right)\dd\eta^{2}+2 B_{i}\dd x^{i}\dd\eta
    +\left(\gamma_{ij}+h_{ij}\right)\dd x^{i}\dd x^{j}\right].
\end{equation}

$B_{i}$ and $h_{ij}$ can be further conveniently decomposed as
\begin{eqnarray}\label{tens-decomp}
B_{i} & = & \partial_{i}B+ \bar{B}_{i}\,, \\
h_{ij} & \equiv & 2C\left(\gamma_{ij}+\frac{\sigma_{ij}}{\HH} \right)+2\partial_{i}\partial_{j}E+2\partial_{(i}E_{j)}+2 E_{ij}\,,
\end{eqnarray}
with
\begin{equation}
 \partial_i \bar{B}^{i}=0=\partial_i E^i, \quad
 E_i^i=0=\partial_iE^{ij}.
\end{equation}
Note that this decomposition of $h_{ij}$ involves the shear. This judicious choice is justified, a
posteriori, by the simplicity of the transformation properties of the perturbation variables, as
we shall now see.

Let us consider an active transformation of the coordinate system defined by a vector field $\xi$.
The coordinates of any point change according to
\begin{equation}\label{coord-trans}
 x^{\mu}\rightarrow\tilde{x}^{\mu}=x^{\mu}-\xi^{\mu}\left(x^\nu\right)
\end{equation}
so that the spacetime metric transforms as
\begin{equation}
 g_{\mu\nu}\rightarrow g_{\mu\nu}+{\cal L}_{\xi}g_{\mu\nu}\,,
\end{equation}
where ${\cal L}_{\xi}g_{\mu\nu}$ is the Lie derivative of $g_{\mu\nu}$ along $\xi$. At first order
in the perturbations, we decompose the metric as $g_{\mu\nu}=\bar g_{\mu\nu} + \delta g_{\mu\nu}$
and it follows that
\begin{equation}\label{lie}
 \delta g_{\mu\nu}\rightarrow \delta g_{\mu\nu}+{\cal L}_{\xi}\bar g_{\mu\nu}.
\end{equation}
The vector field $\xi$ is now decomposed into a scalar and vector part as
\begin{eqnarray}\label{xi_dec}
 &&\xi^0 = T\left(x^{i},\eta\right)\,,\qquad
 \xi^i = \partial^iL(x^{j},\eta)+ L^i\left(x^{j},\eta\right)\,,
\end{eqnarray}
with $\partial_iL^i=0$. With the use of the expressions~(\ref{delta-g}), we deduce that the perturbations
of the metric transform as (in Fourier space)
\begin{eqnarray}
 &&A\rightarrow A + T'+\mathcal{H}T\label{gS1}\\
 &&B\rightarrow B -T+\frac{\left(k^{2}L\right)'}{k^{2}}\\
 &&C\rightarrow C + \mathcal{H}T \\
 &&E\rightarrow E + L\,,\label{gS4}
\end{eqnarray}
for the scalar variables, and as
\begin{eqnarray}
 \bar{B}_i &\rightarrow& \bar{B}_{i} + \gamma_{ij} (L^{j})'-2\mathrm{i}k^j
 \sigma_{lj}P^{l}_{\,\,i} L \label{gV1}\\
 E_i &\rightarrow& E_{i} + L_{i}\,,\label{gV2}
\end{eqnarray}
for the vector variables. We also obtain that the tensor modes are readily gauge invariant,
\begin{equation}
 E_{ij}\rightarrow E_{ij}\,.
\end{equation}
Had we not included the shear in the decompostion~(\ref{tens-decomp}), this would not be the case.

Let us also note that the transformation rule of the vector modes is different from the one
derived in Ref.~\cite{TomitaDen} where the non-commutativity between the projection and the time
evolution has been neglected.

From the gauge transformations~(\ref{gS1}-\ref{gS4}), we
can construct a set of gauge invariant variables for the scalar sector. Only two degrees of
freedom remain, the two other being absorbed by the scalar part of the gauge transformation. We define the
two gravitational potentials
\begin{eqnarray}
  \Phi & \equiv & A+\frac{1}{a}\left\{
  a\left[B-\frac{\left(k^{2}E\right)'}{k^{2}}\right]\right\}'\,,\\
 \Psi & \equiv & -C-\mathcal{H}\left[B-\frac{\left(k^{2}E\right)'}{k^{2}}\right]\,.
\end{eqnarray}
From the gauge transformations~(\ref{gV1}-\ref{gV2}), we deduce that a gauge
invariant vector perturbation is given by
\begin{equation}
 \Phi_{i} \equiv \bar{B}_{i}-\gamma_{ij}\left(E^{j} \right)' +2\mathrm{i}k^j
 \sigma_{lj}P^{l}_{\,\,i} E,
\end{equation}

It is obvious from these expressions that when $\gamma_{ij}$ is time-independent, that is when
$\sigma_{ij}=0$, these variables reduce to the standard Bardeen variables defined in the
Friedmann-Lema\^{\i}tre case. By analogy, we define the Newtonian gauge by the conditions
\begin{equation}
 B=\bar{B}^i=E=0\,,
\end{equation}
so that
\begin{equation}
  A=\Phi\,,\qquad C=-\Psi\,,\qquad \Phi^i=-(E^i)'\,,
\end{equation}
the latter condition being equivalent to $\Phi_i=-E_i'+2\sigma_{ij}E^j$.

\subsubsection{Gauge invariant variables for the matter}

We focus our analysis on the scalar field case, which is the most relevant for the study of
inflation. Under a gauge transformation of the form~(\ref{coord-trans}), it transforms as
$\varphi\rightarrow\varphi+\pounds_\xi\varphi$. At first order in the perturbations, we get
\begin{equation}
 \delta\varphi\rightarrow\delta\varphi + \pounds_\xi\bar\varphi,
\end{equation}
that is
\begin{equation}
 \delta\varphi\rightarrow\delta\varphi + \pounds_\xi\bar\varphi=\delta\varphi+
\varphi' T\,,
\end{equation}
with use of Eq.~(\ref{xi_dec}). Thus, we can define the two gauge invariant variables
\begin{equation}
 Q \equiv \delta\varphi - \frac{C}{\HH}\varphi'
\end{equation}
and
\begin{equation}
 \chi \equiv \delta\varphi + \left[B-\frac{\left(k^{2}E\right)'}{k^2}\right]\varphi'\,.
\end{equation}
They are related by
\begin{equation}
 Q = \chi + \frac{\Psi}{\HH}\varphi'\,.
\end{equation}

\section{Perturbations equations}\label{sec4}

Once the gauge invariant variables have been defined, we can derive their equations of evolution.
The mode decomposition will require a decomposition of the shear tensor in a basis adapted to each
wave-number. We start by defining this decomposition and then we derive the perturbed Klein-Gordon
and Einstein equations.

\subsection{Decomposition of the shear tensor}

The shear $\sigma_{ij}$ is a symmetric tracefree tensor and, as such, has 5 independent components.
In the coordinates system~(\ref{metric2}-\ref{metricdec}), it was expressed in terms of only two
independent functions of time $\beta_i(\eta)$. The 3 remaining degrees of freedom are related to
the 3 Euler angles needed to shift to a general coordinate system.

\subsubsection{Components of the shear}

As mentioned before, when working out the perturbations in Fourier space, it would be fruitful to decompose
the shear in a local basis adapted to the mode we are considering. The shear, being a
symmetric trace-free tensor, can be
decomposed on the basis $\{\hat k_i, e^1_i, e^2_j\}$ as
\begin{equation}\label{decshear}
 \sigma_{ij} = \frac{3}{2}\left(\hat k_i\hat k_j-\frac{1}{3}\gamma_{ij}\right)\spar
 + 2\sum_{a=1,2}\sva \,\hat k_{(i}e^a_{j)}
 + \sum_{\lambda=+,\times}\stl\,\varepsilon^\lambda_{ij}.
\end{equation}
This decomposition involves 5 independent components of the shear in a basis adapted to the
wavenumber $k_i$.  We must stress however that $(\spar,\sva,\stl)$ must not be interpreted as the Fourier
components of the shear, even if they explicitely depend on $k_i$. This dependence arises from the
local anisotropy of space.

Using Eq.~(\ref{decshear}), it can be easily worked out that
$$
 \sigma_{ij}\gamma^{ij} = 0\,,
$$
$$
 \sigma_{ij}\hat k^i = \spar\hat k_j + \sum_a\sva e^a_j\,,
 \qquad
 \sigma_{ij}\hat k^i\hat k^j = \spar\,,
$$
and
$$
 \sigma_{ij}\varepsilon_\lambda^{ij}=\stl\,,\qquad
 \sigma_{ij}\hat k^i e^j_a = \sva\,.
$$
The scalar shear is explicitely given by
\begin{equation}
 \sigma^2 = \sigma_{ij}\sigma^{ij}
          = \frac{3}{2}\spar^2 + 2\sum_a\sva^2 + \sum_\lambda\stl^2\,,
\end{equation}
which is independent of $k_i$. We emphasize that the local positivity of the energy density of
matter implies that $\sigma^2/6 < \HH^2$ and thus
\begin{equation}\label{e:ssurH}
 \frac{1}{2}\spar \leq \frac{1}{\sqrt{6}}\sigma < \HH\,.
\end{equation}
This, in turn, implies that
\begin{equation}\label{e:ssurH2}
 {\spar}<2\HH\,,
\end{equation}
a property that shall turn to be very useful in the following of our discussion. Analogously, we
have that
\begin{equation}\label{e:ssurH3}
 {\stl}<\sqrt{6}\HH\,.
\end{equation}

The following derivations will involve the contraction of the shear with the polarization vectors,
\begin{equation}
 \sigma_{ij} e^i_a e^j_b = -\frac{1}{2}\spar\delta_{ab}
         + \sum_\lambda \stl {\cal M}_{ab}^\lambda\,,
\end{equation}
from which we deduce that
\begin{equation}
 \sigma_{ij} P^{ij} = -\spar\,.
\end{equation}
It will also involve the contraction of the shear with the polarization tensors,
\begin{equation}
 \sigma_{il}\, \varepsilon_\lambda^{lj}=
 -\frac{1}{2}\spar\varepsilon_i^{j\lambda} +\sum_a\sva
 \hat{k}_i e_l^a\varepsilon_\lambda^{lj}+\sum_{\mu}\sigma_{_{\rm T}\mu}
 \varepsilon_{il}^{\mu}\varepsilon_\lambda^{lj}\,,
\end{equation}
which implies that
\begin{equation}
\sigma_{il}\varepsilon^{lj}_{\lambda}P^{i}_{\,\,j}
  =\stl\,,
 \qquad
  \sigma_{il}\varepsilon^{lj}_{\lambda}\varepsilon^{i
  \mu}_{\,\,j}=-\frac{1}{2} \spar \delta^{\lambda}_{\mu}\,.
\end{equation}
To finish, we will make use of the following expression
\begin{equation}
 e_{l}^{b}e_{a}^{j}\,\sigma_{jm}\,\varepsilon_{\lambda}^{lm}  =
-\frac{1}{2}\spar\mathcal{M}_{ab}^{\lambda}+\frac{1}{2}\delta_{b}^{a}\sigma_{_{\rm T}\lambda}+
\frac{1}{2}{\cal N}_{ab}\left(\sigma_{_{\rm T}+} \delta^{\times}_{\lambda}-\sigma_{_{\rm T}\times}
\delta^{+}_{\lambda}   \right)\,.
\end{equation}

\subsubsection{Time evolution of the components of the shear}

In the previous paragraph we detailed the definition of the components of the shear in a basis adapted to the
wave mode $k_i$. The time evolution of these modes is easily obtained from Eq.~(\ref{e:1.14})
\begin{eqnarray}
 &&\spar'+2\HH\spar = -2\sum_a\sigma_{_{\rm V}a}^2\,, \label{background_sigma}\\
 &&\sva' +2\HH\sva = \frac{3}{2}\sva\spar
  -\sum_{b,\lambda}\svb\stl\mathcal{M}_{ab}^\lambda\,, \\
 &&\stl'+2\HH\stl = 2\sum_{a,b}\mathcal{M}_{ab}^\lambda\sva\svb \label{background_sigma3}\,,
\end{eqnarray}
where the matrix $\mathcal{M}_{ab}^\lambda$ is defined in Eq.~(\ref{defMab}).

These equations allow us to derive some important constraints on the rate of change of $\spar$ and
$\stl$. Since Eq.~(\ref{background_sigma}) implies that
\begin{equation}
 \left|\frac{1}{a^2}\left(a^2\spar\right)'\right|=2\sum_a\sigma_{_{\rm V}a}^2<\sigma^2<6\HH^2\,,
\end{equation}
we can conclude that
\begin{equation}\label{e:sparprimsurH}
 \left|\frac{1}{a^2}\left(a^2\spar\right)'\right|<6\HH^2\,.
\end{equation}
Identically, Eq.~(\ref{background_sigma3}) implies that
$$
 \left|\frac{1}{a^2}\left(a^2\stl\right)'\right|= 2\sum_{a,b}\mathcal{M}_{ab}^\lambda\sva\svb
 <\frac{\sigma^2}{\sqrt{2}},
$$
so that
\begin{equation}\label{e:sparprimsurH2}
 \left|\frac{1}{a^2}\left(a^2\stl\right)'\right|<3\sqrt{2}\HH^2\,.
\end{equation}
The two relations~(\ref{e:sparprimsurH}-\ref{e:sparprimsurH2}) will be important at the end of our
analysis.

\subsection{Klein-Gordon equation}

The Klein-Gordon equation, $\Box\varphi=V_{\varphi}$, can be rewritten under the form
\begin{equation}
 g^{\mu\nu}\nabla_\mu\partial_\nu\varphi=V_{\varphi}(\varphi)\,.
\end{equation}
When expanded at first order in the perturbations, the r.h.s. is trivially given by
$V_{\varphi}(\bar\varphi)+V_{\varphi\varphi}(\bar\varphi)\chi$. It follows that the Klein-Gordon
equation at first order in the perturbations is then obtained to be
\begin{eqnarray}\label{e:kg}
 \chi'' +2\HH\chi' -\gamma^{ij}\partial_i\partial_j\chi + a^2V_{\varphi\varphi}\chi =
   2(\varphi''+2\HH\varphi')\Phi +\varphi'(\Phi'+3\Psi')\,,
\end{eqnarray}
where $V_{\varphi\varphi}$ is the second derivative of the potential with respect to the scalar
field. Surprisingly, it has the same form as in the Friedmann-Lema\^{\i}tre case. This can be understood
if we remind that the d'Alembertian can be expressed as $\Box \varphi=
\partial_{\nu}\left[\sqrt{-g}g^{\mu\nu}\partial_{\mu} \varphi \right]/\sqrt{-g}$, and if we realize
that $\sqrt{-g}$ does not involve the shear. Thus, at first order in the perturbations, the only
place where the shear $\sigma_{ij}$ could appear would be associated with $\delta g^{ij}$. But
then it would multiply $\dbi \bar{\varphi}$ which vanishes. Consequently the Klein-Gordon equation
is not modified. This result is not specific to the scalar field case as the conservation equation
in the fluid case is also the same as for a Friedmann-Lema\^{\i}tre spacetime, indeed only as long
as the anisotropic stress vanishes [see Eq.~(\ref{e:1+3_cons})].

\subsection{Einstein equations}

The procedure to obtain the mode decomposition of the Einstein equations is somehow simple. We
start from the general perturbed equation $\delta G^\mu_\nu=\kappa \delta T^\mu_\nu$ with the
expressions~(\ref{dT00}-\ref{dTij}) and~(\ref{dG00}-\ref{dGij}) respectively for the stress-energy
tensor and the Einstein tensor and we then project them, as described in \S~\ref{sec_mode_dec}.

Special care must however be taken. In the Friedmann-Lema\^{\i}tre case, the projections on the
scalar, vector and tensor modes commute with the time evolution. This no more the case in a
Bianchi~$I$ universe, as explained in \S ~\ref{sec_time_pol}. Let us take an example and consider
the extraction of the vector part of an equation involving a term of the form
$(\Phi^i)'+\HH\Phi^i$. We project this equation on the polarization tensor $e^a_i$ to get
$$
 e^a_i\left[(\Phi^i)'+\HH\Phi^i\right] =
    (e^a_i\Phi^i)' - \Phi^i(e^a_i)'+\HH\Phi_a.
$$
We then use Eq.~(\ref{e:eaprim}) to rewrite $\Phi^i(e^a_i)'$, and we develop the shear in the basis
adapted to the mode $k^i$. This implies, in particular, that
contrary to the Friedmann-Lema\^{\i}tre case, the scalar, vector and tensor modes will be coupled.

This being said, the extraction of the mode decomposition of the Einstein equation is a lengthy
but straightforward computation that we carry in the Newtonian gauge. It reduces to (1) Fourier
transforming the Einstein equations, (2) projecting them on the modes, (3) commuting the projection
operators and the time evolution in order to extract the evolution of the
polarizations and (4) finally expressing the decomposition of the shear.

\subsubsection{Scalar modes}

There are 4 scalar Einstein equations. The first is obtained from $\delta G^0_0
=\kappa \delta
T^0_0$ and gives
\begin{eqnarray}\label{e:scal1}
 &&\!\!\!\!\!\!\!\!\!\!\!k^2 \Psi + 3\HH( \Psi'+\HH\Phi) -
        \frac{\kappa}{2}\left(\varphi^{\prime2}\Phi-\varphi'\chi'-V_\varphi a^2\chi\right) =
        \nonumber\\
 &&\!\!\!
        \frac{1}{2}\sigma^2\left[X- 3\Psi\right]
        +\frac{1}{2}\frac{k^2}{\HH}\spar\Psi
        -\frac{1}{2}k^2 \sum_a \tilde \sigma_{_{\rm V}a} \Phi_a
        -\frac{1}{2}\sum_\lambda [\stl E_\lambda' + (\stl'+2\HH \stl)E_\lambda],
\end{eqnarray}
where we have defined the extremely useful variable~\cite{pubook}
\begin{equation}\label{e:defX}
 X \equiv \Phi+\Psi+\left(\frac{\Psi}{\HH}\right)',
\end{equation}
and the quantity
\begin{equation}
 \sva\equiv\mathrm{i} k\tilde \sigma_{_{\rm V}a}\,.
\end{equation}
As an example, the only tricky term which appears when deriving this equation is $\sigma^i_j(E^j_i)'$,
which is obtained from
$$
 \sigma^i_j(E^j_i)' = (\sigma^i_jE^j_i)'-E^j_i(\sigma^i_j)'
                    =  \sum(\stl E_\lambda)' - E^j_i (-2\HH\sigma^i_j)
                    = \sum (\stl E_\lambda)' + 2\HH \stl E_\lambda\,,
$$
where we have used Eq.~(\ref{e:1.14}) to compute $(\sigma^i_j)'$. We will not detail these steps
in the following.

The second equation is obtained from $k^i\delta G^0_i = \kappa k^i\delta T^0_i$. We find
\begin{eqnarray}\label{e:scal2}
 \Psi'+\HH\Phi - \frac{\kappa}{2} \varphi'\chi = -\frac{1}{2\HH}\sigma^2\Psi
 +\frac{1}{2}\spar X
 +\frac{1}{2}\sum_\lambda \stl E_\lambda\,.
\end{eqnarray}
The two remaining equations are obtained from
$$
 \delta^i_j\delta G^j_i=\kappa\delta^i_j \delta T^j_i,\qquad
 \left(\hat k^i\hat k_j-\frac{1}{3}\delta^i_j\right)\delta G^j_i =
 \kappa \left(\hat k^i \hat k_j-\frac{1}{3}\delta^i_j\right)\delta T^j_i
$$
and take the form
\begin{eqnarray}
 && \Psi'' + 2 \HH \Psi'+ \HH \Phi'+ (2 \HH'+ \HH^2)\Phi - \frac{1}{3}k^2
 (\Phi-\Psi) + \frac{\kappa}{2}  \left[ \varphi^{\prime2} \Phi- \varphi' \chi' + V_{\varphi} a^2
  \chi \right] = \nonumber\\
 &&\qquad\qquad
 -\frac{1}{2}\sigma^2\left[X- 3\Psi\right]
 +\frac{1}{6}\frac{k^2}{\HH}\spar\Psi
 +\frac{1}{2} k^2 \sum_a \tilde \sigma_{_{\rm V}a} \Phi_a\nonumber\\
 &&\qquad\qquad
 +\frac{1}{2}\sum_\lambda [\stl E_\lambda' + (\stl'+2\HH \stl)E_\lambda]\,,\label{e:scal3}\\
 && \frac{2}{3}k^2(\Phi-\Psi)=
  \spar\left[X'-\frac{k^2 \Psi}{3\HH}\right]
 +4k^2\sum_{\lambda,a,b}{\cal M}_{ab}^\lambda \tilde{\sigma}_{_{\rm V}}^a
 \tilde{\sigma}_{_{\rm V}}^bE_\lambda - 2k^2\sum_a\tilde{\sigma}_{_{\rm V}a}\Phi_a\,.\label{e:scal4}
\end{eqnarray}
It can be checked that indeed Eqs.~(\ref{e:scal1}, \ref{e:scal2}-\ref{e:scal4}) reduces to their
well-known Friedmannian form when the shear vanishes.

\subsubsection{Vector modes}

The two vector equations are obtained from
$$
 e^i_a \delta G^{0}_{i}=0,\qquad k_i e^j_a\delta G^i_j=0\,.
$$
They respectively give
\begin{eqnarray}\label{e:vec1}
 \Phi_a  &=&
      -2\tilde{\sigma}_{_{\rm V}a} X
      + 4 \sum_{b,\lambda} {\cal M}_{ab}^\lambda \tilde\sigma_{_{\rm V}b}E_\lambda\,,
\end{eqnarray}
and
\begin{eqnarray}\label{e:vec2}
  \Phi'_a +2\HH\Phi_a
  - \frac{5}{2}\spar\Phi_a+ \sum_{b \lambda}\mathcal{M}_{ab}^{\lambda}\sigma_{_{\rm T}\lambda}\Phi_{b}
 &=&  -2\tilde{\sigma}_{_{\rm V}a}X'
 + 4\sum_{b,\lambda}\mathcal{M}_{ab}^{\lambda}\tilde{\sigma}_{_{\rm V}b}{E'}_{\lambda} \nonumber\\
&& + 4 \sum_{b \lambda} {\cal N}_{ab} \tilde{\sigma}_{_{\rm V}b}  \left(\stplus
\delta^{\times}_{\lambda}
  -\stcross \delta^{+}_{\lambda} \right)E_{\lambda}\,,
\end{eqnarray}
where the matrix ${\cal N}_{ab}$ is defined in Eq.~(\ref{defNab}). It can be shown that
Eq.~(\ref{e:vec2}) results from the time derivative of Eq.~(\ref{e:vec1}) once
Eqs.~(\ref{background_sigma}-\ref{background_sigma3}) are used to express the time derivatives of
the shear. This a consequence of the Bianchi identities.

\subsubsection{Tensor modes}

The equation of evolution of the tensor modes is obtained from $\varepsilon_i^{j\lambda}\delta
G^i_j=0$. To simplify, we shall use the shorthand notation $(1-\lambda)$ for the opposite
polarization of $\lambda$, i.e. it means that if $\lambda=+$, then $(1-\lambda)=\times$, and
vice-versa. With the use of Eq.~(\ref{e:bbb}), we obtain
\begin{eqnarray}\label{e:tens1}
E_{\lambda}''+2\HH E_{\lambda}'+k^{2}E_{\lambda} & = & \sigma_{_{\rm T}\lambda}
 \left[k^{2}\left(\frac{\Psi}{\HH}\right)+X'\right]
 + 2 k^2\sum_{a,b}\mathcal{M}_{ab}^{\lambda}\tilde{\sigma}_{_{\rm V}a}\Phi_{b}
\nonumber\\
&& -2k^2\sum_{a}\tilde{\sigma}_{_{\rm V}a}^{2}E_{\lambda} - 2  \stcross \stplus E_{(1-\lambda)} +
2 \sigma_{T (1-\lambda)}^2E_{\lambda}\,.
\end{eqnarray}

It can be shown that in the long wavelength limit, the former equations~(\ref{e:kg},
\ref{e:scal1}-\ref{e:scal4}, \ref{e:vec1}-\ref{e:vec2}, \ref{e:tens1}) are equivalent to the ones
obtained in a more general gradient expansion of Einstein equations on large scales~\cite{cdl}.

\section{Reduced equations and Mukhanov-Sasaki variables}\label{sec5}

The previous equations~(\ref{e:kg}, \ref{e:scal1}-\ref{e:scal4}, \ref{e:vec1}-\ref{e:vec2},
\ref{e:tens1}) form a coupled set of equations for the scalar, vector and tensor modes. In a
Friedmann-Lema\^{\i}tre spacetime, the three kind of perturbations decouple and one can
arbitrarily set one of the contributions to zero to focus on a given type of mode. This is no
more possible here, and in particular, it is not possible to neglect the vector modes. Their
contribution, as we shall see, is in fact central to get the correct set of reduced equations.

First, we introduce the Mukhanov-Sasaki variables~\cite{MSvar} for scalar and tensor modes as
\begin{equation}\label{e:defms}
 v\equiv aQ\,,\qquad
 \sqrt{\kappa}\mu_\lambda \equiv a E_\lambda,
\end{equation}
exactly in the same way as in a Friedmann-Lema\^{\i}tre spacetime. These three variables were
shown to be the canonical degrees of freedom that shall be quantized during inflation
when a Friedmann-Lema\^{\i}tre universe is assumed~\cite{mbf}.

\subsection{Scalar modes}

Let us introduce these variables in our analysis and start by focusing on the scalar modes. First,
we note that Eq.~(\ref{e:scal2}) can be recast under the more compact form
\begin{equation}\label{e:temp2}
 (2 \HH - \spar) X = \frac{\kappa}{a} \varphi'v + \sum_{\lambda} \stl E_{\lambda}\,.
\end{equation}
If we now combine Eq.~(\ref{e:scal1}) with Eq.~(\ref{e:scal3}), replace the vector mode by its
expression~(\ref{e:vec1}) and use the background equations~(\ref{e:a0}), we obtain
\begin{equation}\label{e:temp1}
 \HH X'+ 2(\HH'+ 2 \HH^2)X + \kappa a V_{\varphi} v + k^2 \Psi = \frac{k^2}{3}(\Phi -\Psi) + \frac{2}{3}
 \frac{k^2}{\HH} \spar \Psi\,.
\end{equation}
Now, using Eq.~(\ref{e:scal4}) to simplify the r.h.s., and again replacing the vector mode by its
expression~(\ref{e:vec1}), we get
\begin{equation}\label{magic_equation}
 \left(2 \HH -\spar\right)\left(X'+ \frac{k^2}{\HH}\Psi \right)+
     4 \kappa a^2 V X+ 2 \kappa a V_\varphi v =
   4 k^2\left( \sum_{a} \tilde{\sigma}_{V a}^2X - \sum_{a,b,\lambda}{\cal M}^{\lambda}_{ab}\tilde{\sigma}_{V
  a}\tilde{\sigma}_{V b} E_{\lambda}\right)\,.
\end{equation}
Then, forcing $Q$ in the Klein-Gordon equation~(\ref{e:kg}), using also its background version, we
obtain
\begin{equation}
 Q''+ 2 \HH Q + k^2 Q + a^2 V_{\varphi\varphi} Q + 2 a^2 V_{\varphi} X
 - \varphi'\left( X'+ \frac{k^2}{\HH}\Psi\right)=0\,.
\end{equation}
Now, we can replace the last term by using Eq.~(\ref{magic_equation}) and the next to last by
using Eq.~(\ref{e:temp2}) to get
\begin{eqnarray}
 &&Q''+ 2 \HH Q + k^2 Q + a^2 V_{\varphi\varphi} Q + 2 a^2 V_\varphi X =\nonumber\\
 &&\qquad \frac{\varphi'}{(2 \HH -\spar)}\left[ 4 k^2\left(
\sum_{a} \tilde{\sigma}_{V a}^2X - \sum_{a,b,\lambda}{\cal M}^{\lambda}_{ab}\tilde{\sigma}_{V
  a}\tilde{\sigma}_{V b} E_{\lambda}\right)  - 4 \kappa a^2 V X- 2 \kappa a V'v\right]\,.
\end{eqnarray}
Introducing the definitions~(\ref{e:defms}), we obtain, after some algebra which requires in
particular Eqs~(\ref{background_sigma}-\ref{background_sigma3}) to express terms such as
$\sum_{a,b}{\cal M}^{\lambda}_{ab}\tilde{\sigma}_{V a}\tilde{\sigma}_{V b}$,
\begin{eqnarray}\label{eqforv}
 v''+\left(k^{2}-\frac{a''}{a}+a^{2}V_{,\varphi\varphi}\right)v & = &
\frac{1}{a^2}\left(\frac{2a^2\varphi^{\prime2}}{2\HH-\spar}\right)^{\prime}\kappa v
+\sum_{\nu}\frac{1}{a^2}\left(\frac{2a^2\varphi'\sigma_{_{\rm T}\nu}}{2\HH-\spar}
\right)'\sqrt{\kappa}\mu_{\nu}\,.
\end{eqnarray}
This equation is the first central result of this section.

\subsection{Tensor modes}

The scalar contribution of the tensor equation~(\ref{e:tens1}) is exactly given by the
relation~(\ref{magic_equation}), so that it reduces, after replacing the vector mode by its
expression~(\ref{e:vec1}), to
\begin{eqnarray}\label{eqformu}
\mu_{\lambda}''+\left(k^{2}-\frac{a''}{a}\right)\mu_{\lambda} & = & -2\mu_{\left(1-\lambda\right)}
\sigma_{_{\rm T}+}\sigma_{_{\rm T}\times}+2\mu_{\lambda}\sigma_{_{\rm T} (1-\lambda)}^{2}
+\frac{1}{a^2}\left(\frac{2a^2\varphi'\sigma_{_{\rm T}\lambda}}{2\HH-\spar}
\right)'\sqrt{\kappa}v\nonumber\\
 && +\sum_{\nu}\frac{1}{a^2}\left(\frac{2a^2\sigma_{_{\rm T}\nu} \sigma_{_{\rm
T}\lambda}}{2\HH-\spar}\right)'\mu_{\nu}+\frac{\left(a^2\spar\right)'}{a^2}\mu_{\lambda}
\end{eqnarray}
This equation is the second central result of this section.

\subsection{Summary}

We have reduced the perturbation equations to a set of three coupled equations for the variables
$v$ and $\mu_\lambda$ defined in Eq.~(\ref{e:defms}). If we define two new functions $z_{\rm s}$
and $z_\lambda$ by
\begin{eqnarray}
 \frac{z_{\rm s}''}{z_{\rm s}}(\eta,k_i) &\equiv& \frac{a''}{a}-a^{2}V_{,\varphi\varphi}
 + \frac{1}{a^2}\left(\frac{2a^2\kappa
     \varphi^{\prime2}}{2\HH-\spar}\right)^{\prime}\nonumber\\
 \frac{z_{\lambda}''}{z_\lambda}(\eta,k_i) &\equiv& \frac{a''}{a}
 +2\sigma_{_{\rm T} (1-\lambda)}^{2}+\frac{1}{a^2}\left(a^2\spar\right)'+\frac{1}{a^2}\left(\frac{2a^2\sigma_{_{\rm
T}\lambda}^2}{2\HH-\spar}\right)',
\end{eqnarray}
the system reduces to
\begin{eqnarray}\label{eqforv2}
  v''+\left(k^{2}-\frac{z_{\rm s}''}{z_{\rm s}}\right)v & = &
  \sum_{\nu}\frac{1}{a^2}\left(\frac{2a^2\varphi'\sigma_{_{\rm T}\nu}}{2\HH-\spar}
\right)'\sqrt{\kappa}\mu_{\nu}\,,\\
\mu_{\lambda}''+\left(k^{2}-\frac{z_{\lambda}''}{z_\lambda}\right)\mu_{\lambda} & = &
  \frac{1}{a^2}\left(\frac{2a^2\varphi'\sigma_{_{\rm T}\lambda}}{2\HH-\spar}
\right)'\sqrt{\kappa}v \nonumber\\
&& +\left[\frac{1}{a^2}\left(\frac{2a^2 \stcross
    \stplus}{2\HH-\spar}\right)'-2 \stcross \stplus \right] \mu_{(1-\lambda)}.\label{eqforv3}
\end{eqnarray}
Formally, it can be rewritten as
\begin{equation}
 V'' + k^2 V + \Omega V = \Upsilon V \,,
\end{equation}
where $V\equiv(v,\mu_+,\mu_\times)$. The matrices $\Omega$ and $\Upsilon$ are defined by
\begin{equation}
 V''+
 \left(\begin{array}{lll}
 k^{2}-\frac{z_{\rm s}''}{z_{\rm s}} &0&0\\
 0& k^{2}-\frac{z_{+}''}{z_+}&0\\
 0& 0& k^{2}-\frac{z_{\times}''}{z_\times}
 \end{array}\right)V
 =
 \left(\begin{array}{lll}
 0 &\aleph_+&\aleph_\times\\
 \aleph_+&0&\beth\\
 \aleph_\times& \beth& 0
 \end{array}\right)V\,,
\end{equation}
and the functions $\aleph_\lambda(\eta,k_i)$ and $\beth(\eta,k_i)$ can be read on
Eqs.~(\ref{eqforv2}-\ref{eqforv3}). This is one of the central results of our study.

When the shear vanishes, these equations decouple and we recover the usual
equations~\cite{mbf} for the variables $v$ and $\mu_\lambda$ so that we only have three physical
degrees of freedom. Now, the anisotropy of space is at the origin of some interesting effects.
First the functions $z_{\rm s}$ and $z_\lambda$ are not functions of time only. They depend on
$k_i$ explicitely through the components of the decomposition of the shear. Second, the two types
of modes are coupled through a non-diagonal mass term. The mass term and the evolution operator
cannot be diagonalized at the same time so that we expect the equivalent of a see-saw mechanism.
The importance of the vector modes, that cannot be neglected, has to be emphasized again. Had we
neglected them, the mass term would not be correct.

\subsection{Sub-Hubble limit}

Let us consider the behaviour of the mass term appearing in Eqs.~(\ref{eqforv2}-\ref{eqforv3}) in
the sub-Hubble limit in which $k/\HH\gg1$. We introduce the two slow-roll parameters as
\begin{equation}
 \epsilon\equiv3\frac{\varphi^{\prime2}}{\varphi^{\prime2}+2a^2V}\,,
 \qquad
 \delta\equiv 1-\frac{\varphi''}{\HH\varphi'}\,,
\end{equation}
in terms of which the Friedmann equations take the form
\begin{equation}
 \HH^2=\frac{\kappa}{3-\epsilon}Va^2+\frac{1}{6}\sigma^2\,,
 \qquad
 (3-\delta)\HH\varphi'+V_\varphi a^2=0\,,
\end{equation}
and
$$
\HH'=(1-\epsilon)\HH^2+\left(\frac{\epsilon-3}{6}\right)\sigma^2
$$

We now focus on the behaviour of the functions $\aleph_\lambda$, $\beth$, $z_{\rm s}''/z_{\rm s}$ and
$z_\lambda''/z_\lambda$ in the sub-Hubble regime. We define $x\equiv\sigma/\sqrt{6}\HH$ and use the fact that,
since  $\spar/2 \leq \sigma/\sqrt{6}$ [see Eq.~(\ref{e:ssurH2})], there exists $\alpha<1$ such that
$0 \leq x < \alpha $ due to the positive energy condition [see Eq.~(\ref{e:ssurH})]. Starting from the
definition~(\ref{eqforv2}) for $\aleph_{\lambda}$ we have
\begin{equation}\label{aleph_bound1}
 |\aleph_\lambda| < \left|\frac{1}{a^2}\left(a^2\sigma_{_{\rm T}\lambda}\right)'\right|
 \times  \left|\frac{2\sqrt{\kappa}\varphi'}{2\HH-\spar}\right|
 +2|\sigma_{_{\rm T}\lambda}|
 \times
 \left|\left(\frac{\sqrt\kappa\varphi'}{2\HH-\spar} \right)'\right|\,.
\end{equation}
Now, the property~(\ref{e:sparprimsurH2}) implies that the first term of the right hand side of the inequality
is smaller than
$$
 3\sqrt{2}\HH^2\left|\frac{2 \sqrt{\kappa}\varphi'}{2\HH-\spar}\right|\,.
$$
Then,
\begin{equation}
 \left|\frac{2\sqrt{\kappa}\varphi'}{2\HH-\spar}\right|
 = \sqrt{2 \epsilon} \frac{\sqrt{\HH^2 -\frac{\sigma^2}{6}}}{\HH -\frac{\spar}{2}} \leq
 \sqrt{2 \epsilon}\sqrt{\frac{1+x}{1-x}}.
\end{equation}
Now, since $x$ varies in the range $0 \leq x < \alpha$, we deduce that $\sqrt{(1+x)/(1-x)} \leq
\sqrt{(1+\alpha)/(1-\alpha)}$. Eq.~(\ref{e:ssurH3}) then implies that the second term of the
inequality~(\ref{aleph_bound1}) is smaller than
$$
 2\sqrt{6}\HH
 \times
 \left|\left(\frac{\sqrt{\kappa}\varphi'}{2\HH-\spar} \right)'\right|\,.
$$
Then, the absolute value is bounded by
\begin{eqnarray}
 \left|\left(\frac{\sqrt{\kappa}\varphi'}{2\HH-\spar} \right)'\right|
 &<& \left|\frac{\sqrt{\kappa}\varphi''}{2\HH-\spar}\right| +
    \left|\frac{\sqrt{\kappa}\varphi'}{2\HH-\spar}\right|\left|\frac{2\HH'-\spar'}{2\HH-\spar}\right|. \nonumber
\end{eqnarray}
Using the fact that Eq.~(\ref{e:ssurH2}) implies $|\spar '| < 10 \HH^2$ ($|\spar'| < 6 \HH^2 + |2 \HH \spar|$), we
obtain that
\begin{eqnarray}
 \left|\left(\frac{\sqrt{\kappa}\varphi'}{2\HH-\spar} \right)'\right|
 &\leq& \frac{\sqrt{\epsilon} \HH}{\sqrt{2}} \frac{\sqrt{1+ \alpha}}{\sqrt{1- \alpha}} \left[(1-\delta)
 +\frac{2(1-\epsilon)+\left(1-\frac{\epsilon}{3}\right)\sqrt{6} + 10}{2(1-\alpha)} \right].\nonumber
\end{eqnarray}
Gathering all these terms, we thus conclude that
\begin{equation}
|\aleph_\lambda| < \sqrt{\epsilon} \HH^2 \sqrt{\frac{1+\alpha}{1- \alpha}}  \left[6 + 2
\sqrt{3}\left( (1-\delta)
 +\frac{2(1-\epsilon)+\left(1-\frac{\epsilon}{3}\right)\sqrt{6} + 10}{2(1-\alpha)}\right) \right].
\end{equation}
To summarize, we have shown that
\begin{equation}
 |\aleph_\lambda|< Z \HH^2\,,
\end{equation}
where $Z$ is a finite constant. This constant can in principle be quite large since $\alpha$ can
be arbitrarily close to unity in the worst case of an empty universe. A large $Z$ also corresponds
to a very ellipsoidal Hubble radius, and this explains why the short wavelength limit has to be
taken much smaller than the average Hubble radius.

The same reasoning can be applied for $|\beth|$, $|z_{\rm s}''/z_{\rm s}|$ and $|z_\lambda''/z_\lambda|$.
Thus, it follows that on sub-Hubble scales the three physical degrees of freedom decouple and behave
as harmonic oscillators,
\begin{equation}
 V'' + k^2 V =0 \,.
\end{equation}

\section{Perturbation of the action}\label{sec6}

In order to construct canonical quantization variables and to properly normalize the amplitude of
their quantum fluctuations, one needs to derive the action for the cosmological perturbations. We
will now demonstrate that the previous equations~(\ref{eqforv2}-\ref{eqforv3}) can be obtained
from the expansion of the action, written in the ADM formalism~\cite{aadm}, at second order.
Another simpler route would have been to infer the action from the equations of motion, which is
always possible up to an overall factor, that could then be fixed by considering a simple limiting
case. Still, we prefer to work out the action at second order since it provides a check of the
previous computations.

\subsection{ADM formalism}

In the ADM formalism, we expand the metric as
\begin{equation}\label{adm-metric}
 \dd s^{2}=-\left(N^{2}-N_{i}N^{i}\right)\dd t^{2}+2N_{i}\dd x^{i}\dd t+g_{ij}\dd x^{i}\dd
 x^{j}\,,
\end{equation}
and the Einstein-Hilbert action for a minimally coupled scalar field, takes the
form
\begin{eqnarray}\label{adm-action}
 S&=&\frac{1}{2 \kappa}\int \dd t\dd^{3}x\sqrt{-g}\left[NR^{\left(3\right)}+N\left(K_{ij}K^{ij}-K^{2}\right)
-\kappa N \left(g^{ij}\partial_{i}\varphi\partial_{j}\varphi+2V\left(\varphi\right)\right)\right.\nonumber\\
&& \left.\qquad\qquad\qquad\qquad
+\kappa N^{-1} \left(\dot{\varphi}-N^{i}\partial_{i}\varphi\right)^{2}\right]\,,
\end{eqnarray}
where $R^{(3)}$ is the Ricci scalar constructed with the metric $g_{ij}$ and $K_{ij}$ is the
extrinsic curvature, defined as
\begin{equation}
 K_{ij}\equiv\frac{N^{-1}}{2}\left(\dot{g}_{ij}-2\nabla_{(i}N_{j)}\right),\quad K=K^i_i\,.
\end{equation}
Every spatial index is now manipulated with the metric $g_{ij}$. The ADM metric is designed in
such a way that the constraints arising from the Einstein equations can be immediately derived
from the action. Varying Eq.~(\ref{adm-action}) with respect to the lapse $N$ and the shift
$N_{i}$, we get the Hamiltonian and momentum constraints, respectively
\begin{eqnarray}
&&R^{\left(3\right)}-\left(K_{ij}K^{ij}-K^{2}\right)-2V- \kappa g^{ij}\partial_{i}\varphi\partial_{j}
\varphi+N^{-2} \kappa \left(\dot{\varphi}-N^{i}\partial_{i}\varphi\right)^{2}  = 0\,,\\
&&\nabla_{j}\left(K_{i}^{j}-K\delta_{i}^{j}\right)-N^{-1}\kappa \left(\dot{\varphi}
-N^{j}\partial_{j}\varphi\right)\partial_{i}\varphi  =  0\,.
\end{eqnarray}
Comparing the form~(\ref{dmet1}) of the metric in Newtonian gauge with
Eq.~(\ref{adm-metric}), we conclude that the lapse $N$ and the shift $N_i$ are given by
\begin{equation}
  N^{2}=\left(1+2\Phi\right)\,,\qquad N_{i}=0
\end{equation}
and that the metric $g_{ij}$ is
\begin{equation}
 g_{ij}=a^{2}\left[\gamma_{ij}-2\Psi\left(\gamma_{ij}+\frac{\hat{\sigma}_{ij}}{H}\right)+2\partial_{(i
}E_{j)}+2E_{ij}\right]\,.
\end{equation}
It follows that the Hamiltonian and momentum constraints reduce, at first order, to
\begin{eqnarray}
&&\frac{2}{a^{2}}\Delta\Psi-\frac{1}{a^{2}}\hat{\sigma}^{ij}\partial_{i}\partial_{j}
\left(\frac{\Psi}{H}\right)-6H\dot{\Psi}+\left(\frac{\Psi}{H}\right)^{\mbox{.}}
\hat{\sigma}^{2}-3\Psi\hat{\sigma}^{2}-\Phi\left(6H^{2}-\hat{\sigma}^{2}-\dot{\varphi}^{2}\right)\nonumber\\
&&\qquad\qquad\qquad
+\frac{1}{a}\hat{\sigma}_{j}^{i}\partial_{i}\Phi^{j}-\hat{\sigma}_{j}^{i}\left(E_{j}^{i}\right)
-\kappa V_\varphi\delta\varphi-\kappa\dot{\varphi}\delta\dot{\varphi}=  0\,,
\end{eqnarray}
and
\begin{equation}\label{momentum_constraint}
\hat{\sigma}^{2}\partial_{i}\left(\frac{\Psi}{H}\right)-\hat{\sigma}_{i}^{j}\partial_{j}
\left[\Phi+\left(\frac{\Psi}{H}\right)^{\mbox{.}}\right]+2\partial_{i}\left(\dot{\Psi}
+H\Phi\right)-\frac{1}{2a}\Delta\Phi_{i}+2\hat{\sigma}^{jl}\partial_{j}E_{il}-\hat{\sigma}^{jl}
\partial_{i}E_{jl}- \kappa \dot{\varphi}\partial_{i}\delta\varphi=0\,,
\end{equation}
respectively. Once Fourier transformed, written in conformal time and
projected along its scalar and vector components, we recover precisely Eqs.~(\ref{e:scal1}) and~(\ref{e:scal2}).

In order to expand the action up to second order in all first order perturbed quantities, we
expand the spatial metric as
$$
g_{ij}=a^{2}\left(\gamma_{ij}+h_{ij}\right)\,.
$$
The inverse metric and its determinant are then given by
$$
g^{ij}=a^{-2} \left(\gamma^{ij}-h^{ij}+h^{il}h_{l}^{j}\right)\,,\qquad
\sqrt{g}=a^{3}\left[1+\frac{1}{2}h+\frac{1}{8}h^{2}- \frac{1}{4}h_{j}^{i}h_{i}^{j}\right]\,,
$$
where
\begin{equation}\label{dec_hij}
h_{ij}=-2\Psi\left(\gamma_{ij}+\hat{\sigma}_{ij}/H\right)+2\partial_{(i}E_{j)}+2E_{ij}.
\end{equation}

\subsection{Action at zeroth and first orders}

The expansions of the action at zeroth and first orders are
\begin{eqnarray}
S_{0} & = & \frac{1}{2 \kappa}\int \dd t\dd^{3}x\left[a^{3}\left(-6H^{2}+\hat{\sigma}^{2}-2\kappa V
+ \kappa \dot{\varphi}^{2}\right)\right]\,,\nonumber\\
 & = & \frac{1}{2 \kappa}\int \dd t\dd^{3}x\left[-4\frac{\dd}{\dd t}\left(a^{3}H\right)\right]\,,\label{e511}\\
S_{1} & = & \frac{1}{2 \kappa}\int \dd
t\dd^{3}xa^{3}\left[R_{1}^{(3)}+\hat{\sigma}^{ij}\dot{h}_{ij}
-2\hat\sigma_{ij}\hat\sigma_{l}^{i}h^{jl}+12H\dot{\Psi}+3\Psi\left(6H^{2}-\hat\sigma^{2}+2\kappa
  V
-\kappa \dot{\varphi}^{2}\right)\right.\nonumber\\
 &  & +\left.\Phi\left(6H^{2}-\hat\sigma^{2}-2 \kappa V- \kappa \dot{\varphi}^{2}\right)
-2 \kappa V_\varphi\delta\varphi+2 \kappa \dot{\varphi}\delta\dot{\varphi}\right]\nonumber\\
 & = & \frac{1}{2 \kappa }\int \dd t\dd^{3}x\left\{ \partial_{i}\left[\partial^{i}\left(4a\Psi\right)
-\partial^{i}\left(\frac{a\Psi}{H}\right)^{\mbox{.}}\right]\right.\nonumber\\
 &  &
\left.+\frac{\dd}{\dd t}\left[\Delta\left(\frac{a\Psi}{H}\right)
+a^{3}\hat\sigma^{ij}h_{ij}+12a^{3}H\Psi+2a^{3} \kappa \dot{\varphi}\delta\varphi\right]\right\}
\label{e512} \,,
\end{eqnarray}
where we use the notation $X_n$ for the $n^{\rm th}$ order term of the quantity $X$ when expanded
in perturbations. Note that we have used the background field equations to go from the first line
to the second line in Eqs.~(\ref{e511}-\ref{e512}). As can be seen, these two terms can be
rewritten in terms of total derivatives. It implies that the only nontrivial term will arise from
the expansion of the action at second order.

\subsection{Action at second order}

A lengthy but straightforward computation shows that the expansion of the action at second order
is
\begin{eqnarray}
S_{2} & = & \frac{1}{2 \kappa}\int \dd t\dd^{3}x\; a^{3}\left[R^{(3)}_2+N_{1}R^{(3)}_1+\frac{1}{2}hR^{(3)}_1
+\mathcal{K}_{2}+\frac{1}{2}h\mathcal{K}_{1}+\frac{1}{8}h^{2}\mathcal{K}_{0}\right.\nonumber\\
 &  & -\frac{1}{4}h_{j}^{i}h_{i}^{j}\mathcal{K}_{0}-N_{1}\mathcal{K}_{1}-\frac{1}{2}N_{1}h\mathcal{K}_{0}
+N_{1}^{2}\mathcal{K}_{0}
+\kappa \Big(-a^{-2}\partial_{i}\delta\varphi\partial^{i}\delta\varphi
- V_{\varphi\varphi}\delta\varphi^{2}\nonumber\\
 &  & -2N_{1} V_{\varphi}\delta\varphi-h  V_{\varphi}\delta\varphi-hN_{1} V-\frac{1}{4}h^{2} V
+\frac{1}{2}h_{j}^{i}h_{i}^{j} V+ \delta\dot{\varphi}^{2}
-2N_{1} \dot{\varphi}\delta\dot{\varphi}\nonumber\\
 &  & \left.+N_{1}^{2} \dot{\varphi}^{2}+h \dot{\varphi}\delta\dot{\varphi}
-\frac{1}{2}hN_{1} \dot{\varphi}^{2}+\frac{1}{8}h^{2} \dot{\varphi}^{2}
-\frac{1}{4}h_{j}^{i}h_{i}^{j} \dot{\varphi}^{2}\Big)\right]\,,
\end{eqnarray}
where
\begin{eqnarray}
a^{2}R^{(3)}_1 & = & 4\left(\Delta-\frac{\hat\sigma^{ij}\partial_{i}\partial_{j}}{2H}\right)\Psi\,,\\
a^{2}R^{(3)}_2 & = & -\partial_{l}h^{lj}\partial_{i}h_{j}^{i}-2h^{jl}\partial_{j}
\partial_{i}h_{l}^{i}-9\partial_{i}\Psi\partial^{i}\Psi-\frac{1}{4}\partial_{l}h^{ij}
\partial^{l}h_{ij}-\frac{1}{2}\partial_{l}h_{ij}\partial^{i}h^{lj}\nonumber\\
 &  & -6\partial_{i}\left(h^{ji}\partial_{j}\Psi\right)+\frac{1}{2}\partial^{i}
\partial_{i}\left(h^{jl}h_{jl}\right)\,,\\
\mathcal{K}_{0} & = & -6H^{2}+\hat\sigma^{2}\,,\\
\mathcal{K}_{1} & = & -2H\dot{h}+\hat\sigma^{ij}\dot{h}_{ij}-2\hat\sigma_{ij}\hat{\sigma}_{l}^{j}h^{li}\,,\\
\mathcal{K}_{2} & = & 2H\dot{h}_{ij}h^{ij}-4H\hat\sigma_{ij}h^{il}h_{l}^{j}-2\hat\sigma_{i}^{l}h^{im}
\dot{h}_{ml}+2\hat\sigma_{ij}\hat\sigma_{l}^{j}h^{im}h_{m}^{l}+\frac{1}{4}\dot{h}^{ij}
\dot{h}_{ij}\,,\nonumber\\
 &  & +\hat\sigma_{ij}\hat\sigma_{lm}h^{im}h^{jl}-\frac{1}{4}\dot{h}^{2}\,.
\end{eqnarray}
The construction of the action at second order shall be pursued in Fourier space, since many
non-local operators appear, such as inverse Laplacian $\Delta^{-1}$ or
$(\sigma^{ij}\dbi\dbj)^{-1}$, when using the constraints. Also, it simplifies the use of the
background equations (\ref{background_sigma}-\ref{background_sigma3}) for the components of the
shear $\spar$, $\sva$ and $\stl$ which were defined in Fourier space. We recall that these
components are {\it not} the Fourier transforms of the shear but its decomposition in a basis
adapted to a given mode $k_i$.

The integral of any $3$-divergence is clearly zero in Fourier space. For instance, let us consider
a typical term like $\partial_l(\Psi\partial^l\Psi)$, then
\begin{equation}
\int \dd\eta \dd^{3}x\,\partial_{l}\left(\Psi\partial^l \Psi\right)= \int \dd\eta
\dd^{3}\mathbf{k}\dd^{3}\mathbf{q}\left[-\mathbf{k}\cdot(\mathbf{k}+\mathbf{q})\Psi_\mathbf{k}\Psi_\mathbf{q}
\right]\delta^{(3)}(\mathbf{k}+\mathbf{q})=0\,.
\end{equation}

Thus, we first express $S_2$ in terms of the Fourier modes and then use conformal time. Next,
$h_{ij}$ is replaced by its expression~(\ref{dec_hij}) in function of the variables $\Psi$, $E_j$
and $E_{ij}$. All terms involving $E_j$ either vanish or have the form $\left(E^j\right)'$, and
thus reduce to $-\Phi^j$ in Newtonian gauge. Then, we decompose $\Phi^j$ and $E_{ij}$ according to
Eqs.~(\ref{dec_vector}) and~(\ref{dec_tensor}). The constraint~(\ref{momentum_constraint}), once
expressed in conformal time and in Fourier space, can be projected onto its scalar and vector
parts in order to obtain the scalar constraint~(\ref{e:temp2}) and the vector
constraint~(\ref{e:vec1}). Then, we replace $\Phi_a$ in function of $X$ and $E_{\lambda}$ using
the vector constraint and substitute $\Phi$ by $X-\Psi-\left({\Psi}/{\HH}\right)'$. We then
eliminate $X$ using the scalar constraint. Finally, we replace $E_{\lambda}$ and $\delta \varphi$
by their expressions in terms of the variables $\mu_\lambda$ and $v$ [see Eq.~(\ref{e:defms})].

After a tedious calculation, that strictly follows the recipe described just above, the action
$S_2$ can be recast under a form that contains only the physical degrees of freedom,
\begin{eqnarray}\label{finalS2}
S_{2} & = & \frac{1}{2 }\int \dd\eta \dd^{3}k\Bigg\{ v'v'^{*}+\left(\frac{{z_s}''}{z_s}-k^{2}\right)vv^{*}
+\sum_{\nu}\frac{1}{a^{2}}\left(\frac{2a^{2}\sqrt{\kappa}\varphi'\sigma_{_{\rm T}\nu}}
{2\HH-\spar}\right)'\!\!\left(v^{*}\mu_{\nu}+v\mu_{\nu}^{*}\right)\\
 &  &
 \sum_{\lambda}\left[{\mu'}_{\lambda}{\mu'}_{\lambda}^{*}+\left(\frac{{z_{\lambda}}''}{z_{\lambda}}
-k^{2}\right)\mu_{\lambda}\mu_{\lambda}^{*}+ \left[-2\stcross \stplus+\frac{1}{a^{2}}\left(\frac{2a^{2} \stcross \stplus}
{2\HH-\spar}\right)'\right]\mu_{(1-\lambda)}\mu_{\lambda}^{*}\right]+{\cal T} \Bigg\}\,\nonumber,
\end{eqnarray}
that is, in a more compact way, as
\begin{eqnarray}
S_{2} & = & \frac{1}{2 }\int \dd\eta \dd^{3}k
 \left( |V'|^2 - k^2|V|^2 + {}^tV(\Omega-\Upsilon)V^* +  {\cal T}\right)\,,
\end{eqnarray}
where ${\cal T}$ is a total derivative which, for the sake of completeness, is explicitely given
in Appendix~\ref{appD}.

It is clear under this form that the variation of this action with respect to the physical degrees
of freedom leads directly to the equations of motion~(\ref{eqforv}) and~(\ref{eqformu}). More
important, it shows that the overall factor is unity. It also follows from this action that the
canonical momentum associated with $v$ and $\mu_{\lambda}$ are $\pi_v=v'^{*}$ and $\pi_\lambda=
{\mu'}_{\lambda}^{*}$.

\section{Discussion and conclusions}\label{sec7}

In this article, we have presented a full and complete analysis of the theory of cosmological
perturbations around a homogeneous but anisotropic background spacetime of the Bianchi~$I$ type.
We have described the scalar-vector-tensor decomposition and the construction of gauge invariant
variables. We have reduced our analysis to a scalar field but it can be easily extended to include
hydrodynamical matter.

After presenting the full set of evolution equations for the gauge invariant variables, we have
shown that the vector modes can be algebraically expressed in terms of scalar and tensor modes,
so that only three physical degrees of
freedom remain, one for the scalar sector and two for the tensor sector. Contrary to the
Friedmann-Lema\^{\i}tre case, the scalar, vector and tensor perturbation equations do not
decorrelate and it was important for the consistency of the computation not to neglect the vector
modes. We have shown that these physical degrees of freedom are the trivial generalization of the
Mukhanov-Sasaki variables that were derived in a flat Friedmann-Lema\^{\i}tre universe.

We have also constructed the action for the cosmological perturbations up to second order and
demonstrated that, after use of the constraints was made, it only contains the physical degrees of
freedom and takes a canonical form. We have also shown that in the sub-Hubble limit the scalar and
tensor degrees of freedom decouple and behave as standard harmonic oscillators. It follows that
one can apply the standard quantization procedure~\cite{mbf} and properly define the normalization
of the amplitude of their quantum fluctuations.

The anisotropy of the underlying space induces two physical effects: (1) the equations of motion
explicitely involve the wave-number $k_i$ and (2) a non-diagonal mass term that describes the
coupling between scalar perturbation and gravitational waves is at the origin of a
scalar-tensor see-saw mechanism.

Since the shear decays as the inverse of the second power of the scale factor, the universe
isotropizes and tends toward a Friedmann-Lema\^{\i}tre spacetime. The modes that exit the Hubble
radius during inflation while the shear is non-negligible will experience the see-saw mechanism
and will have the primordial anisotropy imprinted on their statistical properties. Modes of
smaller wavelength will not reflect the anisotropy. It follows that an early Bianchi~$I$ phase may
be at the origin of a primordial anisotropy of the cosmological perturbations, mainly on large
angular scales. The companion article~\cite{pdu} describes such a scenario of early anisotropic
slow-roll inflation. Since the post-inflationary evolution is well described by a
Friedmann-Lema\^{\i}tre spacetime, observable effects, and in particular those related to the CMB
anomalies we alluded to in the introduction, can be taken into account easily once the initial
conditions are known. This investigation, that we plan to do later, is beyond the scope of the present work.

Our analysis extends and sheds some light on the robustness of the quantization procedure that was
developed under the assumption of a Friedmann-Lema\^{\i}tre background, and thus on the
predictions of the standard inflationary scenario. We emphasize that this work is very
conservative and that no new speculative hypothesis was invoked. Indeed, we are not claiming that
such a primordial anisotropy is needed. On the one hand, it can be used to set stronger
constraints on the primordial shear. On the other hand it can also be a useful example for the
study of second order perturbations, in which a shear appears only at first order and induces a
correlation between scalar and tensor at second order~\cite{Gwcoupling,Maldacena}, and more
generally for the understanding of quantum field theory in curved (cosmological)
spacetimes~\cite{Qaniso}. One may for instance wonder whether this analysis can be further
extended to other Bianchi type or to non-spatially flat spacetimes.

\section*{Acknowlegements}

We thank Nathalie Deruelle for her thorough remarks on an early version of this text, Marco
Peloso, Misao Sasaki and Alberto Vallinotto for enlightning discussions and John Barrow, Lev
Kofman and Slava Mukhanov for communication and for pointing to us complementary references. TSP
thanks the Institute of Astrophysics for hospitality during the duration of this work, and the
Brazilian research agency Fapesp for financial support.


\pagebreak
\appendix

\section{Details on Bianchi~$I$ universes}\label{appA}

\subsection{Geometrical quantities in conformal time}\label{app1}

Starting from the metric~(\ref{metricconforme}) in conformal time, the expressions of the
Christoffel symbols are
\begin{equation}
 \Gamma_{00}^{0}=\mathcal{H}\,,
 \quad
 \Gamma_{ij}^{0}=\mathcal{H}\gamma_{ij}+\sigma_{ij}\,,
 \quad
 \Gamma_{0j}^{i}=\mathcal{H}\delta_{j}^{i}+\sigma^{i}_{j}\,.
\end{equation}
where we have used the definition of the shear to express $\gamma'_{ij}=2\sigma_{ij}$ so that
\begin{equation}
  (\gamma^{ij})'=-2\sigma^{ij}\, ,
\end{equation}
and indeed trivially $(\gamma^i_{j})'=(\delta^i_{j})'=0$.

We deduce that the non-vanishing components of the Ricci tensor are given by
\begin{eqnarray}
 a^2 R_0^0 & = & 3\HH'+ \sigma^2 \\
 a^2 R^i_j & = & \left(\HH'+ 2\HH^{2}\right)\delta_{\,j}^{i}+ 2 \HH \sigma^i_j+
                  (\sigma^i_j)',
\end{eqnarray}
where we recall that $\sigma^2=\sigma_{ij}\sigma^{ij}$. The Ricci scalar is
\begin{equation}
 a^2 R=6 \left(\HH'+\mathcal{H}^{2}\right)+ \sigma^2.
\end{equation}

The non-vanishing components of the Einstein tensor are thus given by
\begin{eqnarray}
 a^2 G^0_0 & = & -3\HH^{2}+\frac{1}{2}\sigma^2\label{appg00} \\
 a^2 G^i_j & = & -\left(2\HH'
  +\HH^{2}+\frac{1}{2}\sigma^2 \right)\delta^i_j + 2 \HH \sigma^i_j + (\sigma^i_j)'
  \label{appgij}.
\end{eqnarray}

For a general fluid with stress-energy tensor of the form
\begin{equation}\label{eq:a8}
 T_{\mu\nu}=  \rho u_\mu u_\nu + P(g_{\mu\nu}+u_\mu u_\nu) + \pi_{\mu\nu},
\end{equation}
where $\rho$ is the energy density, $P$ the isotropic pressure and $\pi_{\mu\nu}$ the anisotropic
stress ($\pi_{\mu\nu} u^\mu=0$ and $\pi_\mu^\mu=0$), it implies that the Einstein equation takes
the form
\begin{eqnarray}
 3 \HH^2 &=& \kappa a^2 \rho + \frac{1}{2}\sigma^2\,,\label{e:fried1}\\
 \HH'&=& -\frac{\kappa a^2}{6}(\rho+3P) -\frac{1}{3}\sigma^2\,,\label{e:fried2}\\
 (\sigma^i_j)' &=& - 2 \HH \sigma^i_j +\kappa a^2\tilde\pi^i_j\,,\label{e:fried3}
\end{eqnarray}
which correspond respectively to the ``$00$''-component and trace and trace-free part of the
``$ij$''-equation. The conservation equation for matter reads
\begin{equation}\label{e:cons1}
 \rho'+ 3 \HH (\rho + P) + \sigma_{ij} \tilde{\pi}^{ij} =0\,,
\end{equation}
where the $ij$-component of $\pi_{\mu\nu}$ has been defined as $a^2 \tilde{\pi}_{ij}$ (so that
$\tilde{\pi}^i_{j}=\gamma^{ik}\tilde{\pi}_{kj}$).

To close this sytem, one needs to specify, as usual, an equation of state for the fluid, that is
an equation $P(\rho)$, but also to provide a description for $\pi_{\mu\nu}$. The latter vanishes
for a perfect fluid and for a scalar field.

\subsection{Bianchi~$I$ universes in cosmic time}\label{appA2}

Starting from the metric~(\ref{metric2}) in cosmic time, the expressions of the Christoffel
symbols are
\begin{equation}
 \hat\Gamma_{ij}^0= a^2\left[H\gamma_{ij}+\frac{1}{2}\dot\gamma_{ij}\right]\,,
 \qquad
 \hat\Gamma_{0j}^i= a^2\left[H\delta^i_j+\frac{1}{2}\gamma^{ik}\dot\gamma_{kj}\right]\, .
\end{equation}
The Einstein equations take the form
\begin{eqnarray}
 3 H^2 &=& \kappa\rho + \frac{1}{2}\hat\sigma^2\,,\label{e:fried1C}\\
 \frac{\ddot a}{a}&=& -\frac{\kappa}{6}(\rho+3P) -\frac{1}{3}\hat\sigma^2\,,\label{e:fried2C}\\
 (\hat\sigma^i_j)^. &=& - 3H \hat\sigma^i_j +\kappa \tilde\pi^i_j\,,\label{e:fried3C}
\end{eqnarray}
and the conservation equation for the matter reads
\begin{equation}\label{e:cons1C}
 \dot\rho+ 3 H (\rho + P) + \hat \sigma_{ij} \tilde{\pi}^{ij} =0\,.
\end{equation}

\subsection{Bianchi~$I$ universes in the $1+3$ formalism}\label{appA3}

The dynamics of Bianchi universes can be discussed in terms of the $1+3$ covariant formalism~(see
e.g. Refs.~\cite{vElst,Dunsby}). This description assumes the existence of a preferred congruence
of worldlines representing the average motion of matter. The central object is the 4-velocity
$u^\mu$ of these worldlines. The symmetries imply that it is orthogonal to the hypersurfaces of
homogeneity,
\begin{equation}
 u^\mu = -\delta^\mu_0, \quad
 u_\mu = \delta_{\mu0}.
\end{equation}
The projection operator on the constant time hypersurfaces is defined as
$$
 \perp_{\mu\nu} = g_{\mu\nu} + u_\mu u_\nu.
$$
Its only non-vanishing components being $\perp_{ij} = a^2(t)\gamma_{ij}(t)$.

The central kinematical quantities arise from the decomposition
\begin{equation}\label{dec31}
 \nabla_\mu u_\nu = -u_\mu\dot u_\nu + \frac{1}{3}\Theta \perp_{\mu\nu} + \Sigma_{\mu\nu}
 +\omega_{\mu\nu}.
\end{equation}
For a Bianchi~$I$ universe, homogeneity implies that $D_\mu f=0$ for all scalar functions (where
the spatial derivative operator is defined as $D_\mu T^\alpha = \perp_\mu^{\mu'}
\perp^\alpha_{\alpha'} \nabla_{\mu'}T^{\alpha'}$ etc.) Since $D_\mu P=0$, the flow is geodesic and
irrotational ($\omega_{\mu\nu}=0$) so that the acceleration also vanishes, $a_\mu=0$, and we are
just left with the expansion, $\Theta$, and the shear $\Sigma_{\mu\nu}$.

It is clear from the form~(\ref{metric2}) that
\begin{equation}
 \Theta = 3H.
\end{equation}
The only non-vanishing components of the shear is expressed simply in terms of the trace-free part
of the Christoffel symbol $\hat\Gamma_{ij}^0$ as
$$
 \Sigma_{ij} = a^2\hat\sigma_{ij}
$$
so that
\begin{equation}
 \Sigma^2=\hat\sigma^2=\sum_{i=1}^3\dot\beta_i^2.
\end{equation}

With the general from~(\ref{eq:a8}) for the stress-energy tensor, we get the conservation equation
\begin{equation}\label{e:1+3_cons}
 \dot\rho + (\rho + P)\Theta + \Sigma_{\mu\nu}\pi^{\mu\nu}=0,
\end{equation}
which reduces to Eq.~(\ref{e:cons1}).

The Raychaudhuri~\cite{vElst} equation simplifies to
\begin{equation}
 \dot\Theta + \frac{1}{3}\Theta^2 = - \Sigma^2 -4\pi G(\rho+3P).
\end{equation}
Since $\Theta=3H$, the r.h.s. is simply $3\ddot a/a=3\HH'/a^2$ so that it reduces to
Eq.~(\ref{e:fried2}).

The Gauss equation takes the form
\begin{eqnarray}
 ^{(3)}R_{\mu\nu} &=& - u^\alpha\nabla_\alpha\Sigma_{\mu\nu}
-\Theta\Sigma_{\mu\nu}+\kappa\pi_{\mu\nu}
+\frac{2}{3}\perp_{\mu\nu}\left(\kappa\rho-\frac{1}{3}\Theta^2+\frac{1}{2}\Sigma^2\right).
\end{eqnarray}
We have to be careful here since $u^\alpha\nabla_\alpha\Sigma_{\mu\nu}$ is
not equal to
$\partial_t\Sigma_{\mu\nu}$. It is given, for the $ij$-component by
$u^\alpha\nabla_\alpha\Sigma_{ij}=(a^2\hat\sigma_{ij})^\cdot-2\hat\Gamma_{0j}^k
a^2\hat\sigma_{ik}=a^2\left[\left(\hat\sigma_{ij}\right)^\cdot-2\hat\sigma_{ik}\hat\sigma^k_j\right]$.

In the particular case of a Bianchi~$I$ spacetime, $^{(3)}R_{\mu\nu}=0$ so that the trace of the
generalized Friedmann equation~\cite{vElst} reduces to
$$
\kappa\rho-\frac{1}{3}\Theta^2+\frac{1}{2}\Sigma^2= {}^{(3)}R.
$$
Shifting to conformal time, this gives Eq.~(\ref{e:fried1}) when $^{(3)}R=0$. The trace-free part
leads to Eq.~(\ref{e:fried3}). Note that this implies that when the anisotropic stress vanishes,
$a^3\Sigma_{\mu\nu}$ is constant for the $u^\alpha\nabla_\alpha$ time derivative but that it
implies that $a^2\sigma^i_j$ is constant in terms of the ordinary conformal time derivative. The
identification of $u^\mu\nabla_\mu$ and $\partial_t$ holds only for scalars (see e.g.
Ref.~\cite{Gwcoupling}).

\subsection{General solution of the background equations}\label{appA4}

It is useful to determine general solutions of the evolution of the background
spacetime~\cite{pubook}. We concentrate on the particular case in which $\pi_{\mu\nu}=0$ (relevant
for scalar fields) and first set
\begin{equation}
 \beta_i = B_i W(t).
\end{equation}
Equations~(\ref{beta2}) and (\ref{SigK}) then imply that
$$
 \left(\sum B_i^2\right)\dot W^2(t) = \frac{{\cal S}^2}{a^6}
 \qquad\hbox{or}\qquad
 \left(\sum B_i^2\right)[W'(\eta)]^2 = \frac{{\cal S}^2}{a^4}
 \, ,
$$
from which we deduce that
\begin{equation}
 W(t)=\int\frac{\dd t}{a^3}
 \qquad\hbox{or}\qquad
 W(\eta)=\int\frac{\dd \eta}{a^2}\, .
\end{equation}
The constraints~(\ref{beta}) and~(\ref{beta2}) imply that the $B_i$ must satisfy
\begin{equation}
 \sum_{i=1}^3 B_i=0,\quad
 \sum_{i=1}^3 B_i^2 = {\cal S}^2\, ,
\end{equation}
which are trivially solved by setting
\begin{equation}
 B_i = \sqrt{\frac{2}{3}}{\cal S}\sin\alpha_i,\qquad
 \alpha_p=\alpha+\frac{2\pi}{3}p,\quad p\in\{1,2,3\}.
\end{equation}
Thus, the general solution is of the form
\begin{equation}
 \beta_i(t) =
\sqrt{\frac{2}{3}} {\cal S} \sin\left(\alpha+\frac{2\pi}{3}i\right)\times W,
\end{equation}
where $a$ is solution of
\begin{equation}
 3H^2=\kappa\rho +\frac{1}{2}\frac{{\cal S}^2}{a^6}.
\end{equation}
Once an equation of state is specified, the conservation equation gives $\rho[a]$ and we can solve
for $a(t)$.

As an example, consider the case of a pure cosmological constant, $V={\rm const.}$ and
$\dot\varphi=0$. The Friedmann equation takes the form
$$
 H^2 =  V_0 \left[ 1 + \left(\frac{a_*}{a}\right)^6 \right],
$$
with $V_0\equiv \kappa V/3$ and $a_*\equiv ({\cal S}/6V_0)^{1/6}$. It can be integrated easily to get
\begin{equation}
 a(t) = a_*\left[\sinh\left(3\sqrt{V_0} t \right) \right]^{1/3}.
\end{equation}
Asymptotically, it behaves as the scale factor of a de Sitter universe, $a\propto\exp(\sqrt{V_0}
t)$ but at early time the shear dominates and $a\propto t^{1/3}$.

\section{Properties of the polarizations}\label{appB}

We summarize here the main properties of the polarization tensors, defined in \S~\ref{sec3_1}.

The time derivative of the polarization tensor is given by
\begin{equation}
 \left(\varepsilon_{ij}^{\lambda}\right)'
             = -(\sigma^{kl}\varepsilon_{kl}^\lambda)\, P_{ij}
                                  -(\sigma^{kl}P_{kl})\, \varepsilon_{ij}^{\lambda}
                                  +4\sigma_{(i}^k\varepsilon_{j)k}^{\lambda}\,,
\end{equation}
or equivalently,
\begin{eqnarray}
 \left(\varepsilon_{j}^{i\lambda}\right)'
    &=&  -(\sigma^{kl}\varepsilon_{kl}^{\lambda}) P^i_j
                                  -(\sigma^{kl}P_{kl}) \varepsilon_j^{i\lambda}
                                  +2\sigma_{j}^k\varepsilon_{k}^{i\lambda}\,.
\end{eqnarray}

In terms of the decomposition~(\ref{decshear}) of the shear tensor, it takes the forms
\begin{equation}
 (\varepsilon_{ij}^{\lambda})' =
     -\stl P_{ij} + \spar \varepsilon_{ij}^{\lambda}+4\sigma_{(i}^k\varepsilon_{j)k}^{\lambda}\,,
\end{equation}
and
\begin{eqnarray}
 (\varepsilon_{j}^{i\lambda})'
      =-\stl P^i_j +\spar \varepsilon_j^{i\lambda}+2\sigma_{j}^k\varepsilon_{k}^{i\lambda}\,.
\end{eqnarray}
and the time derivative of the projection operator $P_{ij}$ is given by
\begin{equation}
 (P^{ij})' = -2 \stplus \varepsilon^{ij}_+ + 2 \spar P^{ij}\,.
\end{equation}

Let us also give the following relations that turn to be useful for the derivation of the
evolution equation of the tensor modes
\begin{eqnarray}
\sigma_{il}\varepsilon^{lj}_{\lambda} \sigma_{jp}\varepsilon^{pi}_{\lambda}&=&\frac{1}{4}\spar^2 +
\frac{1}{2}\left(\sigma_{\lambda}^2-\sigma_{(1-\lambda)}^2\right)\nonumber \\
\sigma_{il}\varepsilon^{lj}_{\lambda}
\sigma_{jp}\varepsilon^{pi}_{(1-\lambda)}&=&\stcross \stplus\nonumber\\
\sigma_{il}\varepsilon^{lj}_{\lambda} \sigma^{ip}\varepsilon_{jp}^{\lambda}&=&\frac{1}{4}\spar^2 +
\frac{1}{2}\left(\stplus^2+\stcross^2\right)+\frac{1}{2}\sum_a \sva^2 \nonumber\\
\sigma_{il}\varepsilon^{lj}_{\lambda}
\sigma^{ip}\varepsilon_{jp}^{(1-\lambda)}&=&0\,.\label{e:bbb}
\end{eqnarray}

\section{Perturbed quantities}\label{appC}

For the sake of completeness, let us give the expression of the Lie derivative~(\ref{lie}) of the
displacement $\xi$~(\ref{xi_dec})
\begin{eqnarray}
 {\cal L}_{\xi}\bar g_{00} & = & -2a^{2}\left(T'+\mathcal{H}T\right)\nonumber \\
 {\cal L}_{\xi}\bar g_{0i} & = & a^{2}\left(\xi_{i}'-\partial_{i}T-2
\sigma_{ji}\xi^{j}\right)\nonumber \\
 {\cal L}_{\xi}\bar  g_{ij} & = &
a^{2}\left[2\partial_{(i}\xi_{j)}+2\mathcal{H}T\gamma_{ij}+ 2 T \sigma_{ij}\right]\,.\label{delta-g}
\end{eqnarray}

The expression of the components of the stress-energy tensor of the scalar field at first order is
\begin{eqnarray}
 a^2\delta T_0^0 &=& \varphi^{\prime2}\Phi-\varphi'\chi'-V_\varphi a^2\chi\,,\label{dT00}\\
 a^2\delta T^0_i &=& -\partial_i\left[\varphi'\chi\right]\,,\label{dT0i}\\
 a^2\delta T_i^j &=& -\delta^i_j\left[
\varphi^{\prime2}\Phi-\varphi'\chi'+V_\varphi a^2\chi\right]\label{dTij}\,.
\end{eqnarray}
Note that they are exactly the same expressions than in a Friedmann-Lema\^{\i}tre spacetime. It
comes from the fact that $\delta g_{ij}$ never appears. Indeed, the $\delta T_{ij}$ etc.
components will be different compared to the Friedmann-Lema\^{\i}tre case.

In Newtonian gauge the Christoffel symbols at first order take the form
\begin{eqnarray}
 \delta\Christoffel{0}{0}{0}  &=& \Phi' \\
 \delta\Christoffel{0}{0}{j}  &=& \partial_{j}\Phi \\
 \delta\Christoffel{0}{i}{j}  &=& \HH h_{ij} + \frac{1}{2}h_{ij}'-2 \HH \Phi
                                   \gamma_{ij} -2 \Phi \hat\sigma_{ij}\\
 \delta\Christoffel{i}{0}{j}  &=& \frac{1}{2} h^{i'}_{\,j}- \hat\sigma_{kj}h^{ki}\\
                        & =&
                         \frac{1}{2}(h^{i}_{\,j})'- \hat\sigma_{kj}h^{ki}
                          + h_{kj}\hat\sigma^{ki}\nonumber\\
 \delta\Christoffel{i}{j}{k}  &=& \frac{1}{2}\gamma^{li}\left( \dbj h_{lk} + \dbk
                              h_{jl}-\dbl h_{jk} \right)\,.
\end{eqnarray}
In Newtonian gauge, the expressions of the components of the Einstein tensor at first order are
\begin{eqnarray}
 a^2 \delta G^0_{\,\,0} &=& -2 \Delta \Psi + 6 \HH \Psi' + 2\sigma^2 \Psi -
                     \left(\frac{\Psi}{\HH}\right)'\sigma^2
                     + \frac{{\sigma}^{ij}}{\HH}\partial_i \partial_j \Psi\nonumber\\
                     &&-\sigma^{ij}\partial_i\Phi_j
                     +  \left(E^i_j\right)'\sigma^j_i
                     + (6 \HH^2 -\sigma^2)\Phi\,,\label{dG00}\\
 a^2 \delta G^0_{\,\,i} &=& -\sigma^2 \frac{\partial_i \Psi}{\HH} +
                      \sigma^j_i\partial_j
                   \left[ \Phi + \Psi + \left(\frac{\Psi}{\HH}\right)'\right]
                   -2\partial_i(\Psi'+\HH\Phi)\nonumber\\
                      &&+\frac{1}{2}\Delta\Phi_i
                      -2{\sigma}^{jk}\partial_j E_{ik} +
                     {\sigma}^{jk}\partial_iE_{jk}\,,\label{dG0i}\\
 a^2 \delta G^i_j &=& \delta^i_j \Big [2 \Psi'' + \left(2 \HH^2+4 \HH'
    \right) \Phi +  \Delta \left(\Phi  -\Psi\right)
 + 2 \HH \Phi' + 4 \HH \Psi' \Big]\nonumber\\
 &&+ \dhi \dbj( \Psi -\Phi) -
 \frac{2}{\HH}{\sigma}_{k}^{\,(i}\partial_{j)}\dhk\Psi
  +\sigma_j^i \left[  - \HH\left(\frac{\Psi'}{\HH^2}\right)' +\left(
    \frac{\HH'}{\HH^2}\right)'\Psi + \frac{\Delta \Psi}{\HH} - \Phi'- \Psi'\right]
    \nonumber\\
  &&
    + \delta^i_j \left[\sigma^2 \left( \Phi + \left(\frac{\Psi}{\HH}\right)'-2
  \Psi\right) + \frac{{\sigma}^{kl}}{\HH}\dbk\dbl \Psi \right]
  \nonumber\\
  &&
+(E^i_{\,\,j})'' +  2 \HH (E^i_{\,\,j})'- \Delta E^i_{\,\,j}
 + 2 \left[\sigma^i_k(E^k_{\,\,j})' - \sigma^{k}_{j}(E^i_{\,\,k})'\right]
 - \left[\left(E^{k}_{\,\,\,l}\right)' \sigma^{l}_{k}\right] \delta^{i}_{\,j}
\nonumber\\
  &&+\delta^i_j\sigma^{kl}\partial_k\Phi_l
   -\gamma^{ik} \left[\partial_{(k}(\Phi_{j)})'+2\HH\partial_{(k}\Phi_{j)}
   -2\sigma^l_{(k}\partial_{|l|}\Phi_{j)} \right]\,.\label{dGij}
\end{eqnarray}

\section{Details concerning the expansion of the action}\label{appD}

The total time derivative $\mathcal{T}$ that appears in Eq.~(\ref{finalS2}) is explicitely given
by
\begin{eqnarray}
{\cal T}&=& \Bigg[- a^2 \sigma^{l}_{i}h_{ml}h^{im}+a^2 h_{ij}h^{ij} \HH +
 \HH (a \delta \varphi)^2 -\frac{(\varphi' v)^2}{2 \HH
      -\spar}- \frac{2 a \varphi' v \sigma^{ij}E_{ij}}{2 \HH -\spar}\nonumber\\
&& \quad - \frac{a^2 \left(\sigma^{ij}E_{ij}\right)^2}{2 \HH -\spar}- \HH \mu_{ij}\mu_{kl}
\gamma^{ik}\gamma^{jl}-4 a^2 {\sigma}^{jk}E_{ik} \Psi \left( \gamma^{i}_{j} +
\frac{\sigma^{i}_{\,j}}{\HH}\right)+ \frac{(2\HH
  -\spar) a^2 k^2 \Psi^2}{\HH^2}\nonumber\\
&& \quad -18 \Psi^2 \HH a^2 + \frac{7 \Psi^2 {\sigma}^2 a^2 }{\HH}+\frac{2
  \Psi^2 a^2 \sigma_i^j\sigma_j^k\sigma_k^i}{\HH^2} -\frac{2 v a \Psi \varphi''}{\HH} -6 \Psi
\varphi'a v \Bigg]'\,.
\end{eqnarray}

\end{document}